\documentclass[10pt,a4paper]{article}
\usepackage[latin1]{inputenc}
\usepackage{amsmath}
\usepackage{amsfonts}
\usepackage{amssymb}
\usepackage{graphics}
\usepackage{psfrag}
\usepackage{epsfig}
\usepackage{pstricks}

\newcommand{\m}[1]{\mathcal{#1}}
\newcommand{\nn}{\nonumber}
\newcommand{\ph}[1]{\phantom{#1}}

\author{Jo\~{a}o Magueijo $^1$, Mat\'ias Rodr\'iguez-V\'azquez $^2$,
Hans  Westman $^2$, Tom  Z\l o\'{s}nik $^1$\\
{\small \it $(1)$ Imperial College Theoretical Physics, Blackett Laboratory, London, SW7 2AZ, United Kingdom},\\
{\small \it $(2)$ Instituto de F\'{i}sica Fundamental, CSIC,
Serrano 113-B, 28006 Madrid, Spain}}

\date{\today}
\title{Cosmological signature change in Cartan Gravity with dynamical symmetry breaking}
\begin{document}
\maketitle

\abstract{We investigate the possibility for classical metric signature change in a straightforward generalization of the first order formulation of gravity, dubbed ``Cartan gravity''. The mathematical structure of this theory mimics the electroweak theory in that the basic ingredients are an $SO(1,4)$ Yang-Mills gauge field $A^{ab}_{\ph{ab}\mu}$ and a symmetry breaking Higgs field $V^{a}$, with no metric or affine structure of spacetime presupposed. However, these structures can be recovered, with the predictions of General Relativity exactly reproduced, whenever the Higgs field breaking the symmetry to $SO(1,3)$ is forced to have a constant (positive)
norm $V^aV_a$. This restriction is usually imposed ``by hand'', but in analogy with the electroweak theory we promote the gravitational Higgs field $V^a$ to a genuine dynamical field, subject to non-trivial equations of motion. Even though we limit ourselves to actions polynomial in these variables, we discover a rich phenomenology. Most notably we derive classical cosmological solutions exhibiting a smooth transition between Euclidean and Lorentzian signature in the four-metric. These solutions are non-singular and arise whenever the $SO(1,4)$ norm of the Higgs field changes sign i.e. the signature of the metric of spacetime is determined dynamically by the gravitational Higgs field. It is possible to find a plethora of such solutions and in some of them this dramatic behaviour is confined to the early universe, with the theory asymptotically tending to Einstein gravity at late times. Curiously the theory can also naturally embody a well-known dark energy model: Peebles-Ratra quintessence. }
%
%
\section{Introduction}
As more and more cosmological data pours in, the question remains open as to the need, or not, for modifications to the theory of General Relativity (see, for example, \cite{Clifton:2011jh,Sotiriou:2008rp,Bean:2006up}). Most modifications of gravity in a cosmological setting begin from the second order metric formulation.  In this paper we explore the cosmological behavior of a straightforward generalization of the first order formulation of gravity called Cartan gravity. The mathematical structure of this theory mirrors in key aspects the spontaneous symmetry breaking models of particle physics. It will be shown that such an approach suggests modifications of gravity which would not have been readily considered within the second order formalism. As we shall demonstrate in this paper, such generalizations exhibit interesting and exotic phenomenology, in particular with regards to the issue of classical signature change in cosmology.

The standard (second-order) description of the gravitational field is provided by Einstein's theory of General Relativity, wherein the gravitational field is described solely in terms of the metric tensor $g_{\mu\nu}$. Up to the Hawking-Gibbons-York boundary term, the dynamics is given by the Einstein-Hilbert action:

\begin{eqnarray}
S_{EH}[g_{\mu\nu}] = \frac{1}{16\pi G} \int  \left(R- 2\Lambda\right)\sqrt{-g}\, d^{4}x. \label{eh}
\end{eqnarray}
When considered alongside the matter content of the standard model of particle physics, this theory has enjoyed considerable success in describing the observed properties of space and time on solar system scales. On larger scales yet (ranging from galactic scales to the largest known scales in the observable universe) its success is more conditional. What seems unambiguously true is that to successfully account for observations on these larger scales it is necessary to introduce an additional gravitating component of the universe, dark matter. Furthermore, even given this additional ingredient there seems to be considerable evidence that yet another new dynamical component is involved in the cosmological history, the dark energy. Whether dark matter and dark energy may be regarded as additional matter fields or symptomatic of shortcomings in General Relativity has been the subject of considerable research (see \cite{Clifton:2011jh} and references therein).

The majority of research into possible modifications to gravity has adopted the metric formalism as a starting point. For example, the addition to the action (\ref{eh}) of a term quadratic in the Ricci scalar  \cite{Starobinsky:1980te} appears to do well as an inflaton surrogate \cite{Ade:2013uln}, capable of generating primordial fluctuations. As an alternative explanation of late-time acceleration, authors have considered the effect of the addition of other curvature invariants \cite{Carroll:2003wy,Koivisto:2006ie,Bloomfield:2012ff,Mueller:2012kb,Gubitosi:2012hu} or new gravitational scalar fields \cite{Mota:2010bs,Koivisto:2012za}.
As an alternative explanation of the effects attributed to dark matter, additional scalar and vector fields and tensor fields in the gravitational sector have been considered \cite{Bekenstein:1984tv,Bekenstein:2004ne,ZlosnikFerreiraStarkman2007,Moffat:2007ju,Milgrom:2009gv,Zhao:2010zze,Blanchet:2011wv,Poplawski:2011xf,Babichev:2011kq}.

The metric formalism, however, is not the only description of gravity that can claim to be `unmodified gravity'. An alternative comes in the form of Einstein-Cartan gravity where the descriptors of the gravitational field are a gauge field for the Lorentz group (i.e. the special orthogonal group $SO(1,3)$) $\omega^{IJ}\equiv \omega^{IJ}_{\ph{IJ}\mu}dx^{\mu}$ and a Lorentz-vector valued one-form $e^{I}\equiv e^{I}_{\mu}dx^{\mu}$ (here  $I,J,...=0,1,2,3$). These fields are, respectively, referred to as the spin-connection and the co-tetrad, and have dynamics described by the following action \footnote{For notational compactness we denote the wedge product $y\wedge z$ between differential forms $y$ and $z$ simply as 
$yz$. For example, if $y$ is a one-form and $z$ is a three-form then we have: $\int yz = \int y \wedge z =   \frac{1}{3!} \int y_{\mu}z_{\nu\sigma\delta} dx^{\mu}\wedge dx^{\nu}\wedge dx^{\sigma}\wedge dx^{\delta} =  \frac{1}{3!}\int \varepsilon^{\mu\nu\sigma\delta}y_{\mu}z_{\nu\sigma\delta}  d^{4}x$
where $\varepsilon^{\mu\nu\sigma\delta}$ is the contravariant Levi-Civita density related to the covariant one $\epsilon_{\mu\nu\rho\sigma}$ as $\varepsilon_{\mu\nu\rho\sigma}=g_{\mu\alpha}g_{\nu\beta}g_{\rho\gamma}g_{\sigma\delta}\varepsilon^{\alpha\beta\gamma\delta}=g\epsilon_{\mu\nu\rho\sigma}$, where $g=det(g_{\mu\nu})$ and $\epsilon_{0123}=\varepsilon^{0123}=+1$.}:

\begin{eqnarray}
S_{PH}[e^{I},\omega^{IJ}] &=&  \int  \frac{1}{32\pi G}\left(\epsilon_{IJKL}\left(e^{I}e^{J}R^{KL}-\frac{\Lambda}{6}e^{I}e^{J}e^{K}e^{L}\right)-\frac{2}{\gamma}e_{I}e_{J}R^{IJ}\right)  \label{pal}
\end{eqnarray}
where $R^{IJ}\equiv d\omega^{IJ}+\omega^{I}_{\ph{I}K}\omega^{KJ}$ . This action is equivalent to the Einstein-Hilbert action (with coincident definitions of $G$ and $\Lambda$) only when $e^{I}_{\mu}$ has an inverse (i.e. there exists a well defined vector field $e^{\mu I}$ which satisfies $e^{\mu I}e^{J}_{\nu}\eta_{IJ}=\delta^{\mu}_{\nu}$). If the spin density generated by fermionic matter (which in turn sources torsion) is zero the term involving $\gamma$ (the Holst term) is a boundary term and so does not then contribute to the dynamics of matter and gravity. Since the ensuing gravitational effect of the spin density is typically very small we recover all the predictions of General Relativity. As an aside we note that one may also construct Lagrangians that are $SO(1,3)$ invariant and non-polynomial in $\omega^{IJ}$ and $e^{I}$. This has been the approach of Poincare gauge theory which, in addition to terms present in the constant $\phi$ limit of (\ref{act1}), contains more general terms in the torsion $T^{I}$. This is made possible by use of the tetrad $e^{\mu}_{I}$ in the Lagrangian \cite{Hehl:1994ue,Gronwald:1995em,Chen:2009at,Ho:2011xf,Baekler:2011jt,Hehl:2013qga}.

Therefore the Einstein-Cartan model is a slight generalization of General Relativity in that it does not presuppose that the metric $g_{\mu\nu}\equiv \eta_{IJ}e^{I}_{\mu}e^{J}_{\nu}$ is invertible and so is expected contain more solutions than General Relativity, even when torsion vanishes. However, importantly, it can serve as a starting point for interesting modifications to gravity that may be very difficult to arrive at if beginning from a purely metric formalism \cite{Toloza:2013wi}. Indeed, this approach is largely unexplored compared to the modified gravity literature that takes a metric view of spacetime. In this paper we will explore the cosmological consequences of one of these modifications: Cartan gravity with dynamical symmetry breaking. 

This paper is structured as follows: In Section \ref{cargrav} we present an introduction to gravity as a gauge theory for the de Sitter group, and explain how comparison to the electroweak model suggests a straightforward generalization by allowing its Higgs field to be a truly dynamical field. We will refer
to such gauge theories as Cartan gravity due to their mathematical ingredients being those of Cartan geometry \cite{Randono:2010cq}. In Section \ref{cargrav1} we introduce the  dynamics of the model and discuss how a General Relativistic limit may be obtained. In Section \ref{sign} we highlight the ability of Cartan gravity to dynamically determine the signature of spacetime, including the possibility of signature change, the main focus of the rest of the paper. In Section \ref{frw} we develop the formalism necessary to examine spatially homogeneous and isotropic cosmologies in Cartan gravity. In Section \ref{a1} we present an exact solution to a sub-case of the general Cartan gravity action which displays classical change of metric signature. In Section \ref{a1a2b2} we examine how this solution is affected by the presence of certain other terms in the action, in particular focusing on the recovery of vacuum General Relativity. In Section \ref{b1b2} we demonstrate that another sub-case of the Cartan gravity is equivalent to a Peebles-Ratra `rolling-quintessence' model. 

\section{Cartan gravity with dynamical symmetry breaking}
\label{cargrav}
Let us consider the basic ingredients of the Einstein-Cartan model. The field $\omega^{IJ}$ is an $SO(1,3)$ gauge field, and as such is one of many known gauge fields in physics (alongside the gauge fields of the standard model of particle physics). The co-tetrad $e^{I}$, taken as a fundamental field, has no analogue within Yang-Mills gauge theory: it possesses a spacetime index like a gauge field but does not transform as a gauge field under local $SO(1,3)$ transformations. However, as  understood by MacDowell and Mansouri \cite{MacDowell:1977jt}, and later elaborated upon by Stelle, West, and Chamseddine \cite{Stelle:1979va,Chamseddine:1977ih}, one can regard gravity as a spontaneously broken Yang-Mills type gauge theory. The idea is to enlarge the gauge group from the six dimensional $SO(1,3)$ to one of the ten dimensional groups $SO(1,4)$, $SO(2,3)$, and $ISO(1,3)$, corresponding respectively to the de Sitter, anti-de Sitter, and the Poincar\'e group. Here we shall restrict attention to the de Sitter group.

Cartan gravity is based upon two objects which admit a crisp geometrical interpretation \cite{Westman:2012xk}: an $SO(1,4)$ gauge field $A^{ab}(x)\equiv A^{ab}_{\ph{ab}\mu}(x)dx^{\mu}$  (where $a,b,...=0,1,2,3,4$) and an $SO(1,4)$-valued Higgs field field $V^{a}(x)$. We then imagine a physical situation where $V^{2}\equiv \eta_{ab}V^{a}V^{b}=\mathrm{const.}$, where $\eta_{ab}=\mathrm{diag}(-1,1,1,1,1)$ is invariant under $SO(1,4)$ gauge transformations. If $V^{2}>0$ then we may locally choose a gauge where $V^{a}=(0,0,0,0,\sqrt{V^{2}})$. The group of $SO(1,4)$ transformations $\Lambda^a_{\ph ab}(x)$ that leaves this form of $V^{a}$ unaltered is simply the Lorentz group $SO(1,3)$. As such we can see that the components of the covariant derivative $D_{\mu}V^{a}\equiv \partial_{\mu}V^{a}+A^{a}_{\ph{a}b\mu}V^{b}$ orthogonal to $V^{a}$ (i.e. $D_{\mu}V^{I}\equiv \partial_{\mu}V^{I}+A^{I}_{\ph{a}4\mu}V^{4}=A^{I}_{\ph{a}4\mu}V^{4}$ with $I=0,\dots,3$) transform as an $SO(1,3)$ vector whilst possessing a spacetime index precisely as the co-tetrad $e^{I}$ does. Additionally, it follows that components of 
the gauge field $A^{IJ}$ transform in the same manner as $\omega^{IJ}$, i.e. as an $SO(1,3)$-valued gauge connection. It can be shown \cite{Stelle:1979va,Pagels:1983pq,Randono:2010cq} that the following $SO(1,4)$ covariant action corresponds precisely to the Einstein-Cartan theory:
\begin{eqnarray}
S_{SW}[A^{ab},V^{a},\lambda] &=&  \int \left(  \alpha \epsilon_{abcde}V^{e}F^{ab}F^{cd} + \lambda (V^{2}- V_{0}^{2})\right) \label{sw}
\end{eqnarray}
where $F^{ab}\equiv  dA^{ab}+ A^{a}_{\ph{a}c}A^{cb}$ and the Lagrange multiplier four-form field $\lambda$ enforces the fixed-norm constraint on $V^{2}=V_{0}^{2}$ so as to have symmetry breaking down to $SO(1,3)$. It may be checked that  $16\pi G=-V_{0}/4\alpha$, $\Lambda=+3/V_{0}^{2}$, and $\gamma= \infty$. 

Some preliminary comments are in order.
First we note that the action (\ref{sw}) only contains the two variables $V^a$ and $A^{ab}$: neither metric or an affine structure of spacetime are presupposed in this formulation of gravity. In fact, it is the presence of the symmetry breaking Higgs field $V^a$ that allows for non-trivial dynamics and actions which are not of topological character. Secondly we note that the construction mirrors that of the electroweak theory. In the electroweak theory we have an $SU(2)\times U(1)$-valued Yang-Mills gauge field $B$ and a symmetry breaking $SU(2)\times U(1)$-valued Higgs field $\Phi$ which serves to break the electroweak symmetry leaving the remnant symmetry $U(1)$ of electromagnetism. We can also note that in both cases the Higgs field possess a single degree of freedom, namely the norm (i.e. $V^2$ or $\Phi^\dagger\Phi$), which is not a gauge degree of freedom and is left untouched under actions of the respective gauge group. However, a glaring dissimilarity between the action (\ref{sw}) and the electroweak theory is that while the Higgs field of the electroweak theory is treated as a genuine dynamical field (so that the gauge independent degree of freedom $\Phi^\dagger\Phi$ is subject to non-trivial equations of motion) the 
Higgs field of Cartan gravity is typically treated as a non-dynamical object subject to a restriction $V^2=\mathrm{const.}$ via a Lagrange multiplier. This appears rather {\em ad hoc} from the perspective of the electroweak theory.

The electroweak theory therefore suggests a natural alternative to (\ref{sw}). Instead of imposing $V^2=\mathrm{const}.$ we should treat $V^a$ as a genuine dynamical field and provide dynamical equations of motion for $V^a$ to dictate its behavior. As such the norm $V^2$ can vary and there is no a  priori reason to expect $V^a$ to be always space-like. If the Einstein-Cartan theory is recovered by fixing the norm of $V^{a}$, a generalisation of the Einstein-Cartan model (i.e. a modification of gravity) will follow from allowing $V^{a}$ to vary freely. However this ``modified gravity'' theory would not be even remotely obvious taking the second order formalism as the starting point.  
\section{Polynomial action and General-Relativistic limit}\label{cargrav1}
It is straightforward to write down the most general de-Sitter invariant action which is polynomial in the variables $\{A^{ab},V^{a}\}$:
\begin{eqnarray}
S[A^{ab},V^{a}] &=& \int \bigg(a_{1}\epsilon_{abcde}V^{e}+ a_{2} V_{a}V_{c}\eta_{bd}+a_{3}\eta_{ac}\eta_{bd}\bigg)F^{ab}F^{cd} \nonumber \\
 &&+\bigg(b_{1} \epsilon_{abcde}V^{e}+ b_{2} V_{a}V_{c}\eta_{bd}+b_{3}\eta_{ac}\eta_{bd}\bigg)DV^{a}DV^{b}F^{cd} \nonumber\\
&&+c_{1}\epsilon_{abcde}V^{e}DV^{a}DV^{b}DV^{c}DV^{d}. \label{ca}
\end{eqnarray}
Though this action may look unfamiliar, we can see that it takes on a familiar form in regimes where $V^{2} = \phi^{2} >0$, where $\phi$ is now a dynamical field i.e. it is freely varied and its behaviour is determined, like that of the other fields, by the equations of motion and Lagrange multiplier fields are absent. When the above inequality is satisfied, the symmetry is broken down to $SO(1,3)$ and we may identify $DV^{I}$ with $e^{I}$ and $A^{IJ}$ with $\omega^{IJ}$. The resulting action, up to boundary terms, takes the following form in an $SO(1,4)$ gauge where $V^{a}=\phi\delta^{a}_{\phantom{a}4}$, with $a=\{I,4\}$:

\begin{eqnarray}\label{4action}
S_{L}[\phi,e^{I},\omega^{IJ}] &=&  \int  \frac{1}{32\pi G(\phi)}\left(\epsilon_{IJKL}\left(e^{I}e^{J}R^{KL}-\frac{\Lambda(\phi)}{6}e^{I}e^{J}e^{K}e^{L}\right)-\frac{2}{\gamma(\phi)}e_{I}e_{J}R^{IJ}\right) \nonumber \\
   && + \bigg({\cal C}_{1}(\phi)\epsilon_{IJKL}R^{IJ}R^{KL}+{\cal C}_{2}(\phi)R_{IJ}R^{IJ}+{\cal C}_{3}(\phi)(T^{I}T_{I}-e_{I}e_{J}R^{IJ})\bigg) \label{act1}
\end{eqnarray}
where 
\begin{eqnarray}\label{funcs}
16\pi G(\phi) &=&  \frac{\phi}{2\left(-2a_{1}+b_1\phi^{2}\right)},  \quad
\Lambda(\phi) =  6\frac{\left(a_{1}-b_{1}\phi^{2}+c_{1}\phi^{4}\right)}{\phi^{2}\left(2a_{1}-b_{1}\phi^{2}\right)} , \nonumber\\
\gamma(\phi) &=& 2\frac{\left(2a_{1}-b_{1}\phi^{2}\right)}{(a_{2}+b_{3})\phi},\quad 
{\cal C}_{1}(\phi) =  a_{1}\phi ,  \quad
{\cal C}_{2} (\phi)=  a_{3},  \nonumber\\
{\cal C}_{3}(\phi) &=& \frac{2a_{3}}{\phi^{2}}+\int^{\phi}\left(\frac{2a_{3}}{\phi'^{4}}+\frac{a_{2}}{\phi'^{2}}+\frac{b_{2}}{2}+\frac{b_{3}}{\phi'^{2}}\right) d\phi'^{2}+a_{2}
\end{eqnarray}
and where $T^{I} \equiv de^{I}+\omega^{I}_{\ph{I}J}e^{J}$ is the torsion. 
Note that $\phi$ appears only algebraically, but in fact this is merely a relic of the first-order formalism. Sub-cases of (\ref{act1}) correspond to scalar-tensor theories when converted into second-order language (see, for instance, \cite{Mercuri:2009zi}). This ``algebraic relic'' is analogous to the fact that $e^{I}$ appears only algebraically in the action (\ref{pal}) but the metric from which it is derived appears in (\ref{eh}) via its first and second derivatives. The reason for this is that the dynamics constrain  $\omega^{IJ}$ to be equal to derivatives of $e^{I}$. Upon inclusion of a $\phi$ dependence on ${\cal C}_{3}$ it can be shown that $\omega^{IJ}$ will additionally depend upon derivatives of $\phi$.
However, if it is ${\cal C}_{1}$ and/or ${\cal C}_{2}$ which contain a dependence on $\phi$, it may be shown that one can no longer solve algebraically for all $\omega^{IJ}$: parts exist that obey their own differential equation of motion. In these theories then, parts of the spin-connection (specifically parts of the `contorsion form')  propagate and represent new degrees of freedom in the gravitational sector.

It is worth noting that the various terms in the action \eqref{4action} have already separately been explored in the literature:
\begin{itemize}
\item If it is only $\gamma$ that depends on $\phi$, then we recover the dynamical Immirzi parameter model of \cite{Taveras:2008yf,TorresGomez:2008fj,Calcagni:2009xz}.
\item If if is only only ${\cal C}_{1}$ that depends on $\phi$, then we recover the scalar-Euler form gravity model of \cite{Toloza:2013wi}. 
\item If it is only ${\cal C}_{2}$ that depends on $\phi$ then, we recover the first-order Chern-Simons modified gravity model of \cite{Alexander:2008wi,Alexander:2009tp}.
\item If it is only ${\cal C}_{3}$ that depends on $\phi$, then we recover the Nieh-Yan gravity model of \cite{Mercuri:2009zi}.
\item It was shown in \cite{Westman:2013mf} that the simple action consisting of only $b_1$ and $b_2$ terms corresponds to the extensively studied Peebles-Ratra rolling quintessence model \cite{Ratra:1987rm}.
\end{itemize}

In the limit of constant $\phi$, the action (\ref{act1}) corresponds to the most general $SO(1,3)$ invariant polynomial action that can be constructed from $e^{I}$ and $\omega^{IJ}$ \cite{Rezende:2009sv}. The now-constant functions $\{G,\Lambda,\gamma,{\cal C}_{i}\}$ admit familiar interpretations:  the number $G$ is Newton's constant; the number $\Lambda$ is the cosmological constant; the number $\gamma$ is the Barbero-Immirzi parameter, and the numbers ${\cal C}_{i}$ are constants multiplying, respectively, the Euler (${\cal C}_{1}$), Pontryagin (${\cal C}_{2}$), and Nieh-Yan (${\cal C}_{3}$) boundary terms. 
General Relativity in its Einstein-Cartan form is therefore exactly  reproduced whenever $V^2= const$. This provides us a clear General-Relativistic limit for the Cartan model, corresponding to $V^2\rightarrow const$. Departures from General relativity in the action \eqref{act1} are therefore encoded entirely in the dependence of any of $\{G,\Lambda,\gamma,{\cal C}_{i}\}$ upon a non-constant $\phi$ (which in turn is controlled by the $\{a_i,b_i,c_i\}$ parameters of the original action). 

In considering the nature of these departures, one may worry about the non-polynomial appearance of the field $\phi$ in (\ref{4action}) and the implications this may have for stability of the theory. This is, however, of course, merely a relic of the use of the `compound' variable $e^{I}$: for example the polynomial term $A^{I}_{\phantom{I}4}A^{4J}$ becomes $-\frac{1}{\phi^{2}}e^{I}e^{J}$. In the following calculations we will instead opt to use variables constructed from $A^{ab}$ and $V^{a}$ such that the Lagrangian remains polynomial\footnote{However, we note that the existence of polynomial Lagrangians and equations of motion in itself cannot guarantee the absence of pathological behaviour. e.g. consider the equation $dx/dt=x^{3}$ whose general solution becomes singular at finite $t$ for positive initial values of $x$ }.

\section{The prospect of signature change in the new theory}\label{sign}
The coupling between the Einstein-Cartan fields $\{e^{I},\omega^{IJ}\}$ and $\phi$ described by (\ref{act1}) allows for a considerable amount of modification to standard gravitation. However, these modifications cover only regimes where the $SO(1,4)$ norm of the Higgs field satisfies $V^{2}=V^aV_a>0$, something which is {\it not} imposed as a constraint. If, for instance, there exist solutions where $V^{2}<0$ over some region of the spacetime manifold, then the remnant symmetry of the theory is not $SO(1,3)$ but instead $SO(4)$, i.e. the four dimensional Euclidean group, and one may utilise a gauge where $V^{a}=\psi \delta^{a}_{0}$, with $a=\{0,{\cal I}\}$ (where ${\cal I},{\cal J},...$ now represent four-dimensional Euclidean indices). Then one may deduce an analog to (\ref{act1}) describing a very general coupling of a scalar field $\psi$ to Euclidean Einstein-Cartan gravitational fields $\omega^{{\cal I}{\cal J}}=A^{{\cal I}{\cal J}}$ and $e^{{\cal I}}=DV^{{\cal I}}$. In the limit $\psi\rightarrow const$ it may be seen that Euclidean Einstein-Cartan gravity with a cosmological constant plus boundary terms are recovered.

Therefore, as $V^{a}$ is now regarded as a genuine dynamical field with its own equations of motion, it is conceivable that there exist solutions where $V^{2}$ changes sign, and thus the signature of spacetime changes, in the sense of the remnant symmetry group  will vary. These solutions do exist and in fact appear naturally. The rest of this paper will be devoted to exhibiting them, and discussing their patterns. In order to better appreciate their significance, it will be  useful to start by discussing the status of the metric signature in the second order formalism of General Relativity and in the Einstein-Cartan model. 

In General Relativity the possibility of classical signature change remains controversial. One may take the view that the field equations alone determine what kind of solutions are allowed. Then, the restriction to globally hyperbolic spacetimes can be regarded as an {\em ad hoc} restriction. Instead, we may  regard solutions with closed time-like curves (e.g. the G\"odel and Kerr solutions) to be physically allowed spacetimes, as they appear naturally as exact solutions of the Einstein field equations. Although this view is controversial, it may equally well be applied to the issue of the signature of spacetime, i.e. how many space- and time-dimensions we have. As a demonstration that signature change is indeed possible within General Relativity one may consider the Einstein field equations sourced by a minimally coupled scalar field in FRW symmetry, and search for cosmological solutions fitting into the ansatz $g_{tt}=f(t)$ where $f(t)< 0$ for $t>t_{0}$ and $f(t)>0$ for $t<t_{0}$ \cite{Ellis:1991st,Dereli:1993pj,GhafooriTabrizi:1999nf}. These solutions do exist, but they are quite distinct from  those about to be shown here. Our solutions are dynamically determined by the evolution equations; instead the sign of $g_{tt}$ in the second order formalism is not determined by the Einstein field equations.

In contrast, the signature of the spacetime metric in four-dimensional Einstein-Cartan theory is unambiguous, and signature change is not possible. To see this consider the $SO(1,3)$ Einstein-Cartan model and recall that the four dimensional metric follows from the relation $g_{\mu\nu}=\eta_{IJ}e^{I}_{\mu}e^{J}_{\nu}$. Due to the signature of the $SO(1,3)$ invariant matrix $\eta_{IJ}=\mathrm{diag}(-1,1,1,1)$ it is impossible to construct a metric $g_{\mu\nu}$ with signature $(+,+,+,+)$ for real $e^{I}_{\mu}$. There are caveats to this argument, found by extending the number of spacetime dimensions (as opposed to the internal symmetry group, as in Cartan theory). Since (\ref{pal}) does not assume invertibility of the matrix $e^{I}_{\mu}$, there may exist solutions where there are regions where the metric has signature $(0,+,+,+)$ or $(-,0,+,+)$, thereby ``obliterating'' one dimension. Therefore apparent signature change would be possible, for example, taking a 5D space with signature $(-,+,+,+,+)$, and transitioning from degenerate solutions of the form $(-,+,+,+,0)$ to those of the form $(0,+,+,+,+)$. Similar transitions via degenerate solutions can be used to implement topology change in the Einstein-Cartan formalism~\cite{Horowitz:1990qb}. Nonetheless it is true that if we restrict ourselves to a fixed number of target space dimensions, then signature change in the Einstein-Cartan formalism  appears forbidden.

By enlarging the internal group to $SO(1,4)$ and then breaking it via a Higgs field valued on this group, the situation is quite distinct from these two cases, as we now show.

\section{FRW Symmetry}
\label{frw}
Let us now consider cosmological solutions. These are solutions that are both homogeneous and isotropic on three-dimensional sub manifolds, i.e. display FRW symmetry. Due to the fact that the basic variables $A^{ab}$ and $V^a$ carry gauge indices ($a,b,\dots$) it is not straightforward to impose FRW symmetry, i.e. we cannot naively require the solutions to satisfy the standard Killing equations ignoring the gauge indices. How this problem is circumvented is explained in detail in Appendix \ref{FRWSYM} where it is shown that the most general functional form in spherical coordinates $(t,r,\theta,\varphi)$ of $A^{ab}$ and $V^a$ satisfying FRW symmetry is
\begin{eqnarray}
V^a&\overset{*}{=}&(\psi(t),0,0,0,\phi(t)) \label{vansatz} \\
A^{ab}&\overset{*}{=}&\left(\begin{array}{ccc}0&B(t)E^j&N(t)E^0\\-B(t)E^i&\omega^{ij}&A(t)E^i\\-N(t)E^0&-A(t)E^j&0\end{array}\right)
\end{eqnarray}
where 
\begin{eqnarray}
E^1=\frac{dr}{K(r)}\quad E^2=rd\theta\quad E^3=r\sin\theta d\varphi\qquad K(r)=\sqrt{1-kr^2}\qquad k=-1,0,+1
\end{eqnarray}
and \cite{Toloza:2013wi}

\begin{eqnarray}
\omega^{0i}=BE^i \quad \omega^{12}=-\frac{K(r)}{r}E^2-CE^3\quad \omega^{13}=-\frac{K(r)}{r}E^3+CE^2\quad \omega^{23}=-\frac{\cot\theta}{r}E^3-C E^1
\end{eqnarray}
and $C=C(t)$. The curvature $F^{ab}$ becomes
\begin{eqnarray}
F^{kl}&=&-\dot C\epsilon^{kl}_{\ph{kl}m}E^0E^m+(k+B^2-A^2-C^2)E^kE^l\\
F^{0j}&=&(\dot B-NA)E^0E^j+BC\epsilon^j_{\ph jmn}E^mE^n\\
F^{j4}&=&(\dot A-NB)E^0E^j+AC\epsilon^j_{\ph jmn}E^mE^n
\end{eqnarray}
where $E^{0}\equiv dt$ and a dot denotes a derivative with respect to $t$.

Note that we have only partially fixed the gauge, i.e. we have imposed $V^i=0$, $i=1,2,3$, but allowed for a non-zero $V^0=\psi(t)$. This is necessary, since requiring $\psi=0$ would unduly exclude a time-like symmetry breaking field $V^a$, and so be unable to cover a signature change event. Given the identification (see Section \ref{cargrav}) of $e^{I}$ with $DV^{I}$ when $V^{2}>0$ and $e^{{\cal I}}=DV^{{\cal I}}$ when $V^{2}<0$ and given the ansatz (\ref{vansatz}) we identify the 3-metric on surfaces of constant $t$ as follows:

\begin{eqnarray}
h_{\mu\nu} \equiv  \delta_{ij}D_{\mu}V^{i}D_{\nu}V^{j}\equiv a(t)^{2}\delta_{ij}E_{\mu}^{i}E_{\nu}^{j}
\end{eqnarray}
where the function $a(t)$ is the scale-factor. To go beyond this in the general case we need one further formal development.

\subsection{Covariant Formalism}
\label{covfor}
It is possible to cover situations in which space-like and time-like $V^a$ fields are present by allowing for a different partial gauge fixing, in which two components of $V^a$ are allowed to be non-vanishing, one space-like, one time-like. 
In using a form for $V^{a}$ with two independent components $(\psi,\phi)$ we retain a residual $SO(1,1)$ gauge freedom. Under such a $SO(1,1)$ gauge transformation the components $\psi$ and $\phi$ transform as an $SO(1,1)$ vector. Therefore we can consider a new object, $SO(1,1)$ vector $V^{A}=(\psi,\phi)$, where the indices $A,B,...$ (i.e. Latin capitals in the first half of the alphabet) can only take two values: 0 and 4 (as opposed to $I,J,...$ used, e.g. in~(\ref{4action}), which run from 0 to 3). Furthermore, by inspection $A^{Ai}=(BE^{i},-AE^{i})$ transforms as a one-form valued in the group $SO(1,1)\times SO(3)$, which we will denote $W^{A}E^{i}$.
Finally it may be checked that $A^{AB}$ transforms as an $SO(1,1)$ gauge field whilst $\omega^{ij}$, as expected, transforms as an $SO(3)$ gauge field. Therefore we may express the  components of the curvature $F^{ab}$ under FRW symmetry in a manifestly $SO(1,1)\times SO(3)$ covariant manner:
\begin{eqnarray}
F^{kl}&=&-\dot C\epsilon^{kl}_{\ph{kl}i}E^0E^i+(k-W^2-C^2)E^kE^l\\
F^{Aj}&=&\m {\cal D}W^AE^0E^j+W^AC\epsilon^j_{\ph jmn}E^mE^n\\
F^{AB}&=&0
\end{eqnarray}
where ${\cal D}$ is the $SO(1,1)$ covariant derivative. Furthermore we have that:

\begin{eqnarray}
DV^{i} &=&  -W^{A}V_{A}E^{i} \\
DV^{A}  &=& {\cal D}V^{A}\, .
\end{eqnarray}
By comparison with the definition of $h_{\mu\nu}$ we may identify the scale factor:

\begin{equation}
a(t)\equiv-W_{A}V^{A}
\end{equation}
which, as expected, is a gauge-invariant quantity. 
\subsection{Metric tensors}\label{metr}
From a Cartan-geometric point of view the metric structure of the manifold is given by
\begin{align}
g_{\mu\nu}=P_{ab}D_\mu V^a D_\nu V^b
\end{align}
where $P_{ab}=\eta_{ab}-\frac{V_aV_b}{V^2}$ is a projector. However, there are other symmetric second rank tensors that can be constructed from the pair $\{A^{ab},V^{a}\}$. This situation is similar to that of scalar-tensor theory in the second-order formalism of gravity i.e. where the fields are a spacetime metric $\mathrm{g}_{\mu\nu}$ and scalar field $\alpha(t)$. There one has the freedom to define a class of other metrics on spacetime via the following \emph{disformal relation} \cite{Koivisto:2012za,Zumalacarregui:2010wj}:

\begin{eqnarray}
\tilde{\mathrm{g}}_{\mu\nu} &\equiv & \tilde{f}_{1}(\alpha) \mathrm{g}_{\mu\nu}- \tilde{f}_{2}(\alpha)\partial_{\mu}\alpha\partial_{\nu}\alpha
\end{eqnarray}
where a choice of functions $\tilde{f}_{1}$ and $\tilde{f}_{2}$ specify the transformation. As $\alpha=\alpha(t)$ we have that the non-`time-time' components of $\tilde{\mathrm{g}}_{\mu\nu} $ and $ \mathrm{g}_{\mu\nu}$ agree up to the time-dependent scaling $\tilde{f}_{1}$. Analogously, consider the following tensor:
\begin{eqnarray}
{\cal G}_{\mu\nu} &\equiv &  D_{\mu}V^{a}D_{\nu}V_{a} =\delta_{ij}D_{\mu}V^{i}D_{\nu}V^{j}  + \eta_{AB}{\cal D}_{\mu}V^{A}{\cal D}_{\nu}V^{B} \label{gee0}
\end{eqnarray}
and a class of tensors

\begin{eqnarray}
\tilde{{\cal G}}_{\mu\nu} &\equiv&\tilde{F}_{1}(V^{2})D_{\mu}V^{a}D_{\nu}V_{a}+\tilde{F}_{2}(V^{2})D_{\mu}V^{2}D_{\nu}V^{2} \label{metrics}
\end{eqnarray}
One particularly important tensor corresponds to the metric $g_{\mu\nu}=\eta_{IJ}e^{I}_{\mu}e^{J}_{\mu}$ whenever $V^{2}\equiv \eta_{ab}V^{a}V^{b}>0$ (viz. equation (\ref{4action})) and is given by the following choice for functions: 

\begin{eqnarray}
g_{\mu\nu} &\equiv &  \left(\eta_{ab}-\frac{V_{a}V_{b}}{V^{2}}\right)D_{\mu}V^{a}D_{\nu}V^{b} \\
  &=& D_{\mu}V^{a}D_{\nu}V_{a}-\frac{1}{4V^{2}}D_{\mu}V^{2}D_{\nu}V^{2} \label{gee1}
\end{eqnarray}
i.e. it is the part of ${\cal G}_{\mu\nu}$ with gradients of $V^{2}$ projected out.

We note that when $V^{2}>0$ the signature of $g_{\mu\nu}$ is $(-,+,+,+)$ and, as may be checked, when $V^{2}<0$ the signature of $g_{\mu\nu}$ is $(+,+,+,+)$. At any moment when $\partial_{t}V^{2}=0$ , all $\tilde{{\cal G}}_{\mu\nu}$ are related by a conformal factor $\tilde{F}_{1}$ and so agree on the metric signature.
Note that $g_{\mu\nu}$ is not necessarily well-defined at $V^{2}=0$.

\section{Classical Signature Change in the simplest case}
\label{a1}

We first examine the case where the action consists of only the $a_{1}$ term and $a_{1}$ does not depend on $V^2$. We will refer to this as the $a_{1}$-action. Note that this action is identical to the action (\ref{sw}) in the absence of the fixed-norm constraint upon $V^{a}$. We now proceed to write this action in terms of the covariant notation of the previous section:

\begin{eqnarray}
S_{a1}&=& \int a_1 \epsilon_{abcde}V^eF^{ab}F^{cd}=\int 4 a_1\epsilon_{AjklB}V^BF^{Aj}F^{kl}\nn\\
&=& \int 4 a_1\epsilon_{jkl}\epsilon_{AB}V^B\left(\m DW^AE^0E^j(k-W^2-C^2)E^kE^l-W^AC\epsilon^j_{\ph jmn}E^mE^n\dot C\epsilon^{kl}_{\ph{kl}i}E^0E^i\right)\nn\\
&=& \int dt4 a_1\bar V_A\left(\m DW^A(k-W^2-C^2)-2\dot CCW^A\right)\int_\Sigma \epsilon_{jkl}E^jE^kE^l
\end{eqnarray}
where we have introduced the notation $\bar{V}_{A}\equiv \epsilon_{AB}V^{B}$. The integration over the spatial hypersurface $\Sigma$ can be carried out trivially and  we can read off the FRW reduced action as
\begin{eqnarray}
S_{a1(FRW)}&=&\int dt4 a_1(\bar V_A\m DW^A(k-W^2-C^2)-2\dot CCW^A\bar V_A)\nn\\
&=&\int dt4 a_1(\bar V_A\m DW^A(k-W^2-C^2)+C^2\m D(W^A\bar V_A))
\end{eqnarray}
Varying with respect to $N$, $W^A$, $V^A$, and $C$ yields:
\begin{eqnarray}
0&=&V_AW^A(k-W^2-C^2)\\
0&=&-\m D\bar V_A(k-W^2-C^2)+\bar V_A\m DW^2-2W_A\bar V_B\m DW^B\\
0&=&\m D\bar W_A(k-W^2-C^2)-2\dot CC\bar W_A\\
0&=&CW^A\m D\bar V_A.
\end{eqnarray}
Recalling the definition of the scale factor $a(t)\equiv -W_{A}V^{A}$, we recognize that the first equation reads $a(k-W^{2}-C^{2})=0$. Generally we do not expect the scale factor to be always zero, so we choose to impose the condition $k-W^2-C^2=0$ in the remaining equations.
We make the following ansatz for $W^{A}$:

\begin{eqnarray}
W^{A} =  \frac{1}{2}\left(k-C^{2}+1\right){\cal W}^{A}+\frac{1}{2}\left(k-C^{2}-1\right)\bar{\cal W}^{A} \label{hansatz}
\end{eqnarray}
where ${\cal W}^{A}{\cal W}_{A}=1$. One can check that this ansatz is indeed consistent with $W^{A}W_{A}=(k-C^{2})$. Because of the unit norm condition on ${\cal W}^{A}$, we can always parameterise it as follows:

\begin{eqnarray}
{\cal W}^{0} = \sinh f(t)    \quad  {\cal W}^{4} = \cosh f(t) \label{param}
\end{eqnarray}
We see from the equations of motion that we must evaluate ${\cal D}W^{A}$. We see that given the ansatz (\ref{hansatz}) and the parameterisation (\ref{param}) we may write this covariant derivative in terms of time derivatives of $C$ and the quantity ${\cal D}{\cal W}^{A} =-(\dot{f}+N)\bar{{\cal W}}^{A}\equiv {\cal N}\bar{{\cal W}}^{A}$. Now we use these results in the equations of motion to yield:

\begin{eqnarray}
0&=&\dot CC\bar W_A\\
0&=&-\left(\m N a-\dot CC\frac{\bar a+a}{k-C^2}\right)W_A-\dot CC\bar V_A\\
0&=&-2C\left(\m N a-\dot CC\frac{\bar a+a}{k-C^2}\right)+2C\dot{\bar a}
\end{eqnarray}
where $\bar{a}\equiv \bar{V}_{A}W^{A}$. Assuming that $W_{A}\neq 0$ and $C\neq 0$, we then have that

\begin{eqnarray}
\dot{C} &=& 0 \\
\dot{\bar{a}} &=& 0 \label{barev}\\
{\cal N} &=& -(\dot{f}+N) =0
\end{eqnarray}
whereas the condition $W^{2}=(k-C^{2})$ may be written in the form:

\begin{eqnarray}
a^{2}-\bar{a}^{2}=(k-C^{2})V^{2}. \label{lapsn}
\end{eqnarray}
It is now important to introduce some notion of proper time. In Section \ref{metr} we considered two tensors $g_{\mu\nu}$ and ${\cal G}_{\mu\nu}$ from which we can extract two different proper times both agreeing with each other in the $V^2=\mathrm{const.}$ regime. Using the above results and the definitions \eqref{gee1} and \eqref{gee0} yields 
 
 \begin{eqnarray}
{\cal G}_{\mu\nu}dx^{\mu}dx^{\nu} &=&  \frac{1}{k-C^{2}}\left(\frac{d\sqrt{\bar{a}^{2}+V^{2}(k-C^{2})}}{dt}\right)^{2}dt^{2}+a^{2}\delta_{ij}E^{i}_{\mu}E^{j}_{\nu} \label{cgmn}
\end{eqnarray}
and
\begin{eqnarray}
g_{\mu\nu}dx^{\mu}dx^{\nu} &=&   -\frac{\bar{a}^{2}\dot{a}^{2}}{(k-C^{2})^{2}V^{2}}dt^{2}+a^{2}\delta_{ij}E^{i}_{\mu}E^{j}_{\nu}. \label{met1} 
\end{eqnarray}
From (\ref{cgmn}) we may define a proper length $T$ according to $G_{\mu\nu}dx^{\mu}dx^{\nu}$:
\begin{eqnarray}
dT^{2} &=& \frac{1}{k-C^{2}}d\sqrt{\bar{a}^{2}+V^{2}(k-C^{2})}^{2}.
\end{eqnarray}
First we consider the case where $k-C^{2}>0$. Recalling the constancy of $\bar{a}$ and $C$, this can readily be integrated to obtain:
\begin{eqnarray}
V^{2} =  -\frac{\bar{a}^{2}}{(k-C^{2})}+\frac{1}{4}(T-T_{0})^{2} \label{vs1}
\end{eqnarray}
where $T_{0}$ is a constant of integration. Furthermore, using equation (\ref{lapsn}) we have

\begin{eqnarray}
a^{2}= \frac{(k-C^{2})}{4}(T-T_{0})^{2}. \label{as1}
\end{eqnarray}
We see then that in terms of $T$, $V^{2}$ is negative between times $T= T_{0} \pm 2\bar{a}/(k-C^{2})^{1/2}$, reaching a minimum value of $-\bar{a}^{2}/(k-C^{2})$ at $T=T_{0}$. At $T_{0}$ we have that $a^{2}=0$. At other times, $V^{2}>0$ is positive and both $V^{2}$ and $a^{2}$ grow without bound.
The solution for the case $k-C^{2}<0$ follows simply from the substitution $V^{2}\rightarrow -V^{2}$ in (\ref{vs1}) and $(k-C^{2})\rightarrow -(k-C^{2})$ in (\ref{as1}).

From (\ref{met1}) we may define an alternative proper length $\tau$ according to $g_{\mu\nu}dx^{\mu}dx^{\nu}$.
From (\ref{met1}) we see that the sign of $g_{tt}$ depends only on the sign of $V^{2}$.
For $k-C^2>0<1-\frac{a^2}{\bar a^2}$ we have
\begin{align}
d\tau=\frac{da}{\sqrt{(k-C^2)(1-\frac{a^2}{\bar a^2})}}
\end{align}
where $\tau$ is the proper-time. Hence we get

\begin{eqnarray}
a  = \bar{a}\sin \left(\frac{\sqrt{(k-C^{2})}}{\bar{a}}(\tau-\tau_{-})\right). \label{fors}
\end{eqnarray}
For $k-C^2>0<\frac{a^2}{\bar a^2}-1$ we get
\begin{align}
d\tau=\frac{da}{\sqrt{(k-C^2)(\frac{a^2}{\bar a^2}-1)}}
\end{align}
which integrated becomes
\begin{eqnarray}
a= \bar{a}\cosh\left({\frac{\sqrt{k-C^{2}}(\tau-\tau_{+})}{\bar{a}}}\right). \label{ford}
\end{eqnarray}
The corresponding $V^{2}(\tau)$ can simply be read off from (\ref{lapsn}). 

Recall the definitions $a=-W_{A}V^{A}$ and $\bar{a}= \epsilon_{AB}W^{A}V^{B}$. For solutions where  $a$ and $\bar{a}$ are non-zero when $V^{2}=0$, the field $V^{A}$ is passing through the null cone from one-sign norm to another rather than actually involving the point $V^{A}=0$. Any solutions involving $V^{A}(T_{0})=0$ are expected to concomitantly have $a(T_{0})=\bar{a}(T_{0})=0$. Any moment where $V^{A}=0$ is a moment where symmetry breaking due to $V^{A}$ is absent. Such initial data would seem to imply from equation (\ref{vs1}) that

\begin{eqnarray}
V^{2}=\frac{\mathrm{sign}(k-C^{2})}{4}(T-T_{0})^{2}   \qquad (\mathrm{if}\,\,\, V^{A}(T_{0})=0). \label{fn}
\end{eqnarray}
An identical solution follows by requiring that instead only $W^{A}(T_{0})=0$ or  $V^{A}(T_{0})=W^{A}(T_{0})=0$.  It can be checked that for these solutions ${\cal G}_{\mu\nu}(T_{0})=0$. 

Note that $\mathrm{sign}({\cal G}_{TT})= \mathrm{sign}(k-C^{2})$ and $\mathrm{sign}(g_{\tau\tau})= \mathrm{sign}(V^{2})$. Therefore, for example, although $V^{2}$ changes sign in the solution described by equations (\ref{vs1}) and (\ref{as1}), the sign of ${\cal G}_{TT}$ is never negative and so signature change according to the metric ${\cal G}_{\mu\nu}$ does not happen. By definition the metric $g_{\mu\nu}$ is directly sensitive to the remnant symmetry of the field equations given by some $V^{2}(x^{\mu})$: $SO(4)$ when $g_{\tau\tau}>0$ and $SO(1,3)$ when $g_{\tau\tau}<0$. We will see in Section \ref{a1a2b2} that there exist solutions for more complicated cases (i.e. cases involving more parameters of the action being non-zero) where every possible metric of the class defined in (\ref{metrics}) `agree' that signature change has taken place.

Clearly then the determinant of whether $V^{2}$ grows unbounded with respect to $T$ or $\tau$ in regimes where it is positive or negative depends only on the sign of the constant of the motion $k-C^{2}$. The case $k-C^{2}=0$ is special. From (\ref{lapsn}) we immediately see that here we have $a^{2}=\bar{a}^{2}$ i.e. a static universe. 
\begin{figure}
\label{betag0}
\centering
\includegraphics[width=8cm]{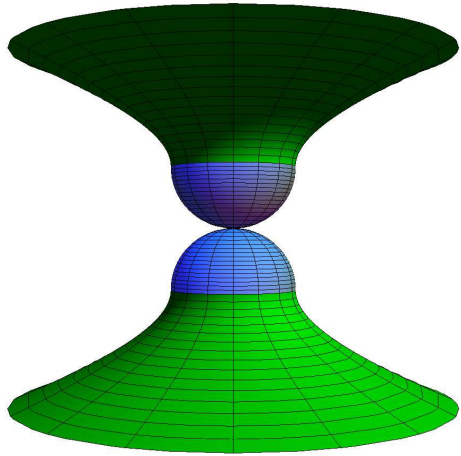} 
\caption{{\small Evolution of the modulus of the scale factor for a solution following from an action described by an $a_{1}$ term for the case $k=1$, $C=0$. $g_{tt}<0$ for green areas and $g_{tt}>0$ for blue areas. Evolution is displayed with respect to a Cartesian coordinate of a flat space (blue region) or spacetime (green region), which one may consider the surface to be piecewise embedded in. As discussed in Section \ref{interp}, care must be taken for the interpretation of the interface between green and blue regions and the meeting of the `south and north pole' within the blue region.}}
\end{figure} 

\subsection{Interpretation of results}
\label{interp}
The above solutions display a number of properties that are unfamiliar from the metric Riemannian perspective. For example, at different moments during the evolution the metric components $g_{\tau\tau}$ or $g_{ij}$ vanish, rendering the spacetime metric non-invertible and degenerate. One may worry that these instances represent singularities of some kind, perhaps signified by the divergence of spacetime scalars which explicitly involve the inverse metric, for instance the Ricci scalar $R\equiv g^{\mu\nu}R_{\mu\nu}$ which requires a well-defined metric inverse $g^{\mu\nu}$. 

To address these concerns we must keep in mind the mathematical framework in which these solutions were obtained. The fundamental field variable in a Cartan-geometric formulation is not the spacetime metric. Rather, the fundamental field variables are $V^a$ and $A^{ab}$ which always appear polynomially in the equations. This should be contrasted to the Einstein field equations in which the metric inverse appears frequently. Thus, within a metric Riemannian formulation the absence of a well-defined metric inverse leads to mathematical difficulties for the differential equations. In contrast, the solution in Fig. \ref{betag0} comes from evolving the fields $A^{ab}$ and $V^{a}$ using equation which are polynomial. When imposing FRW symmetry the partial differential equations reduces to first order ordinary differential equations with respect to a suitable cosmic time parameter. If the solutions are smooth over the entire manifold, thus rendering the polynomial Lagrangian four-form smooth and finite, then it becomes appropriate to view these solutions as non-singular. This condition for acceptability of solutions is more general than the requirement that the metric tensor $g_{\mu\nu}(V,A)$ constructed from the basic dynamical variables $\{A^{ab},V^{a}\}$ must always be invertible. 

Furthermore, even though the scale factor $a$ becomes zero we stress that this has no bearing on what the underlying topology is. Indeed, the equations of motion and their solutions are defined on a manifold with topology either ${\cal R}\times {\cal R}^{3}$ $(k=0$), ${\cal R}\times {\cal H}^{3}$ $(k=-1)$, or ${\cal R}\times {\cal S}^{3}$ $(k=1)$. That the scale factor may vanish at some time does not change this.  

The solutions (\ref{fors}) and (\ref{ford}) are remarkably simple and provide a classical realization of the Hartle-Hawking no-boundary proposal (with one caveat to be discussed below). When $V^{2}<0$ the scale-factor is described by (\ref{fors}) and, choosing the arbitrary constant $\tau_{-}$ to be $0$, yields the following spacetime metric for $|\tau | < \tau_{+} \equiv\pi\bar{a}/(2\sqrt{k-C^{2}})$:

\begin{eqnarray}
g_{\mu\nu}dx^{\mu}dx^{\nu}&=&   d\tau^{2}+  \bar{a}^{2}\sin^{2} \left(\frac{\sqrt{(1-C^{2})}}{\bar{a}}\tau\right) d\Omega_{3}^{2}\\
&=&  \frac{\bar{a}^{2}}{1-C^{2}} \left(d\beta^{2}+(1-C^{2})\sin^{2}\beta d\Omega^{2}_{3}\right)
\end{eqnarray}
where $\beta=\frac{\sqrt{(1-C^{2})}}{\bar{a}}\tau$, $d\Omega_{3}^{2}$ is the metric of the unit three-sphere, and we have used the fact that $k=1$ for this solution. We see that when $C=0$ this is the metric of a four-sphere with radius $\bar{a}$. However, the solution obtained from the Cartan-geometric equations of motion do not `stop' at the south pole. Instead, as shown in Fig. \ref{betag0}, attached to the south pole we find a north pole and the solution extends `past' the south pole. This must be contrasted to the Hartle-Hawking no-boundary proposal in which it would be nonsensical to ask what `happened before' the big bang as this would be like asking `what is south of the the south pole?'. We might be worried about the potentially unhealthy looking `pinch' in the geometry (the moment $\tau=0$), joining a south pole to a north pole. However, within the polynomial Cartan-geometric formulation such a pinch is  more accurately thought of as a `moment' where the scale factor is zero and the spatial metric degenerate over a sub-manifold with the topology of ${\cal S}^{3}$. 

We see that the particular form of a solution (i.e. that the scale factor $a$ becomes zero) does not dictate the underlying topology in a Cartan geometric formulation. Indeed, this example is highly reminiscent of the example considered by Horowitz \cite{Horowitz:1990qb} for the metric $g_{\mu\nu}dx^{\mu}dx^{\nu}=-dt^{2}+t^{2}dx^{2}+dy^{2}+dz^{2}$ where $x,y,z$ are identified with $x+1,y+1,z+1$ respectively. For $y$ and $z$ constant the two-metric is that of a cone when $t\neq 0$. For $t=0$ the metric is non-invertible but in the Einstein-Cartan formulation of gravity the solution yielding this metric is described by fields $e^{I}$ and $\omega^{IJ}$ which are smooth for $-\infty \leq t \leq +\infty$. Again, the polynomial character of the Einstein-Cartan equations of motion removes the necessity of a well-defined metric inverse.

For $0<C<1$, the factor $\sqrt{1-C^{2}}$ may be absorbed into the definition of angular coordinates on the three-sphere, attenuating their range by this factor. This would appear to be a higher dimensional generalization of the `American football' geometry that can be achieved by removing an angular section covering azimuthal angle $\phi_{0}$ from a two-sphere (see for example Figure 1 of \cite{Carroll:2003db}). Treated as a metric geometry, one would usually regard $\beta=0$ as the `location' of a conical singularity; as in the case where $C=0$ though, it seems more accurate in this case to think of this again as the location of a degenerate spatial metric \footnote{Interestingly, this interpretation of degenerate metrics on sub-manifolds of ${\cal R}\times {\cal S}^{3}$ manifolds has been examined in detail for the case where the metric is degenerate not on ${\cal S}^{3}$ sub-manifolds but ${\cal R}\times {\cal S}^{2}$ sub-manifolds i.e. the spatial metric is taken to be degenerate on ${\cal S}^{2}$ sub-manifolds of ${\cal S}^{3}$ and for all-moments of time rather than merely an instant as in this case \cite{Randono:2009gy,Randono:2010cq}}.

When $V^{2}>0$ the scale factor is described by (\ref{ford}) and yields the following spacetime metric for $|\tau| >\tau_{+}$:

\begin{eqnarray}
g_{\mu\nu}dx^{\mu}dx^{\nu}&=&   -d\tau^{2}+  \bar{a}^{2}\cosh^{2} \left(\frac{\sqrt{(1-C^{2})}}{\bar{a}}(\tau-\tau_{+})\right) d\Omega_{3}^{2}\\
&=&  \frac{\bar{a}^{2}}{1-C^{2}}\left(-d\alpha^{2}+(1-C^{2})\cosh^{2}\left(\alpha-\alpha_{+}\right) d\Omega^{2}_{3}\right)
\end{eqnarray}
where $V^{2}(\tau_{+})=0$. This metric corresponds to the metric of de-Sitter spacetime with de-Sitter radius $\bar{a}$ when $C=0$. It may be checked that the curvature tensor $F^{ab}=0$ vanishes here, as one may expect. 

The moment where $V^{2}=0$ marks the transition between Euclidean and Lorentzian regimes. In line with our previous remarks there is nothing pathological here as our basic fields are smooth, continuous, and differentiable at this moment just as at all other points on the manifold. Whether this behaviour persists following the introduction of realistic matter content is quite another issue and we postpone this question for future work. However, we note that geodesics are well-behaved in the geometry of Fig. \ref{betag0} (see e.g.  \cite{Ellis:1991st}). Furthermore, it is encouraging to note that actions for matter may be readily constructed which are polynomial in matter and the fields $\{A^{ab},V^{a}\}$ and free of terms which require invertibility of the metric \cite{Pagels:1983pq,Westman:2012zk}.  Another way to further probe whether signature change and `pinches' causes problems is to study perturbations of $V^a$ and $A^{ab}$ around this background solution.
\section{Adding other terms to the $a_{1}$ action}
\label{a1a2b2}
We have seen in Section \ref{cargrav} that the terms due to $a_{1}$ reduce to the Palatini action plus cosmological constant when $V^{2}$ is constant. However, the solutions of Section \ref{a1} do not dynamically lead to a freezing of $V^{2}$, and therefore an approximate reduction to Einstein gravity. This means that although this model is a good toy-model for signature change in the first order formalism, it cannot be taken as a viable model for our Universe. 
It is therefore necessary to consider what happens if other terms are `switched-on', together with $a_{1}$, in the hope that this may lead to more realistic models. This will also give us some insight into the robustness of the signature change solution of the previous section.
A simple addition is to consider a $b_{2}$ term alongside the $a_{1}$ term of (\ref{ca}). This term explicitly contains gradients of $V^{2}$ and so should be sensitive to the dynamics of $V^{2}$.
In FRW symmetry, the $a_1-b_2$ action takes the form:
\begin{eqnarray}
S_{a1b2}
&=&\int dt4 a_1\left(\bar V\cdot\m DW(k-W^2)+C^2W\cdot \m D\bar V-\chi CV\cdot \m DV (W\cdot V)^2\right)
\end{eqnarray}
where $\chi\equiv b_{2}/8a_{1}$. As in the case where only the $a_{1}$ term is non-zero, variation with respect to $N$ yields the constraint:

\begin{eqnarray}
a(W^{2}-k+C^{2})=0\, .
\end{eqnarray}
Therefore we may again make the ansatz (\ref{hansatz}) expressing $W^{A}$ as a function of $k$, $C^{2}$ and the unit-spacelike vector ${\cal W}^{A}$, which again is parameterised by a function $f(t)$.
After calculation (detailed in Appendix \ref{devab}) it can be shown that the remaining field equations may be cast in the form
\begin{eqnarray}
\dot{(V^{2})} &=&  - \frac{{\cal N}(a-\bar{a})}{\chi C V^{2}}\\
\dot{C} &=&  \frac{{\cal N}(a-\bar{a})\bar{a}}{CV^{2}}\\
\dot{\bar{a}} &=&  -\frac{{\cal N}(a-\bar{a})(a^2-2C^{2}V^2)}{2C^{2}V^{2}}
\end{eqnarray}
where we recall that ${\cal N}\equiv -(\dot{f}+N)$. Additionally we have the constraint provided by the $N$ equation of motion:
\begin{eqnarray}
a^{2}-\bar{a}^{2} = (k-C^{2})V^{2}\, .
\end{eqnarray}
As in the case where only $a_{1}$ is non-zero, we may use the forms of $g_{tt}$ and ${\cal G}_{tt}$ to relate the coordinate $t$ to the proper distances $\tau$ and $T$:
\begin{eqnarray}
g_{tt}  &=&  -{\cal N}^{2}\frac{(k-C^{2})^{2}(a^{2}\chi V^{2}-\bar{a}C)^{2}}{4\chi^{2}a^{2}(a+\bar{a})^{2}C^{4}V^{2}}
\end{eqnarray}
\begin{eqnarray}
{\cal G}_{tt} &=& -{\cal N}^{2}\frac{(k-C^{2})(C(k-C^{2})+a^{2}\chi(a+\bar{a}))(C(C^{2}-k)+a^{2}\chi(a-\bar{a}))}{4\chi^{2} a^{2}(a+\bar{a})^{2}C^{4}}.
\end{eqnarray}
This fully specifies our mathematical problem. However, a further manipulation significantly clarifies the presentation of its solutions.

\subsection{Dimensionless quantities}
In our equations we have only one dimensionful constant, namely $\chi$ which has dimensions of $L^{-3}$. The magnitude $|\chi|$ only serves to rescale the variables in a solution and it is therefore a good idea to eliminate it by introducing dimensionless variables:
\begin{align}
a=\alpha |\chi|^{-1/3}\qquad \bar a=\bar\alpha |\chi|^{-1/3}\qquad C=\m C\qquad V^2=\m V^{2}|\chi|^{-2/3}.
\end{align}
The equations of motion then take the form
\begin{eqnarray}
\dot{\m V}^{2} &=&  - \frac{{\cal N}(\alpha-\bar\alpha)}{\m C {\m V}^{2}} \label{ev1}\\
\dot{\m C} &=&  \frac{{\cal N}(\alpha-\bar{\alpha})\bar\alpha}{\m C{\m V}^{2}}\\
\dot{\bar\alpha} &=&  -\frac{{\cal N}(\alpha-\bar\alpha)(\alpha^2-2\m C^{2}{\m V}^{2})}{2\m C^{2}{\m V}^{2}} \label{ev3}
\end{eqnarray}
whilst the constraint becomes
\begin{equation}
\alpha^2-\bar\alpha^2=(k-\m C^2){\m V}^{2} \label{lapseq}
\end{equation}
and the $tt$ component of the metrics $g_{\mu\nu}$ and $\m G_{\mu\nu}$ turn into
\begin{eqnarray}
|\chi|^{2/3}g_{tt}&=&-\m N^2\frac{(k-\m C^2)^2(\alpha^2{\m V}^{2}-\bar\alpha\m C)^2}{4\alpha^2(\alpha+\bar\alpha)^2{\m C}^4{\cal V}^{2}} \label{gtt}\\
|\chi|^{2/3}{\cal G}_{tt} &=& -{\cal N}^{2}\frac{(k-\m C^{2})(\m C(k-\m C^{2})+\alpha^{2}(\alpha+\bar\alpha))(\m C(\m C^{2}-k)+\alpha^{2}(\alpha-\bar\alpha))}{4\alpha^{2}(\alpha+\bar\alpha)^{2}\m C^{4}}\, .
\end{eqnarray}
These are the variables in terms of which we will explore the space of solutions for our theory.

\subsection{Solutions}

Collectively we have five fields ${\cal N}(t),\alpha(t), \bar{\alpha}(t),{\cal C}(t),{\cal V}^{2}(t)$. We choose a form of ${\cal N}$ such that, via (\ref{gtt}), the coordinate time $t$ coincides with proper-time $\tau$.
The equations (\ref{ev1}) to (\ref{ev3}) are first-order evolution equations and so we need to specify the values of fields at some initial moment $\tau_{0}$ in order to solve them. In choosing initial data $\alpha(\tau_{0})$, ${\cal C}(\tau_{0})$, and ${\cal V}^{2}(\tau_{0})$ we can further obtain a value for $\bar{\alpha}(\tau_{0})$ via the constraint equation (\ref{lapseq}). As this involves taking a square-root, there are two allowable values of $\bar{\alpha}(\tau_{0})$ for each $\{\alpha(\tau_{0}), {\cal C}(\tau_{0}), {\cal V}^{2}(\tau_{0})\}$: $+|\bar{\alpha}(\tau_{0})|$ and $-|\bar{\alpha}(\tau_{0})|$. By inspection of the equations of motion,  evolution from initial data $\{\alpha(\tau_{0}), {\cal C}(\tau_{0}), {\cal V}^{2}(\tau_{0}),-|\bar{\alpha}(\tau_{0})|\}$ is of identical functional form to evolution from initial data $\{-\alpha(\tau_{0}), -{\cal C}(\tau_{0}), {\cal V}^{2}(\tau_{0}),+|\bar{\alpha}(\tau_{0})|\}$. Therefore, in exploring the solution space, it is sufficient to always consider the value $+|\bar{\alpha}(\tau_{0})|$ as the evolution from considering the other square root may be found by simply considering different initial values for the triple  $\{\alpha,{\cal C},{\cal V}^{2}\}$.

We have investigated the properties of this system numerically in detail for the case $k=1$, 
finding the general properties illustrated and enumerated in the parametric plots of solutions $\alpha(\tau)$ and ${\cal V}^{2} (\tau)$ displayed in Figure 2. Figure \ref{3d}, plotting $\{\alpha(\tau), {\cal C}(\tau), {\cal V}^{2}(\tau)\}$,  illustrates further these various cases, 
 showing that the contorsion scalar ${\cal C}(\tau)$ plays a crucial role in the diversity of these solutions. The following qualitatively different types of solution (labelled in Figure 2) may  be identified:

\begin{enumerate}
\item In Case $1$ there is no signature change. From Figure 2 it can be seen that the magnitude of the dimensionless scale factor $\alpha$ tends to $\infty$ asymptotically, reaching a finite minimum value at an intermediate time. This may be interpreted as eternal contraction of the universe, pause of contraction at finite $\alpha$, then an infinite period of expansion i.e. the solutions describe a non-singular bouncing universe with unchanging metric signature.
Cases $1a)$ and $1b)$ represent indistinguishable universes---they differ only by arbitrary choice of orientation of the basis one-forms $E^{i}$. ${\cal V}^{2}$ asymptotically tends to differing constant values of the same sign as proper time tends to $-\infty$ and $+\infty$. By inspection of the action \eqref{4action} we see then that asymptotically we recover Lorentzian general relativity with differing, necessarily positive, values of the cosmological constant $\Lambda$. 

\item In Case 2, ${\cal V}^{2}$ oscillates eternally between positive and negative values as illustrated in Figure 4. As also shown, $\alpha$ also oscillates around $\alpha=0$ eternally, reaching a maximal $|\alpha|$ before returning to $\alpha=0$. Although $\alpha=0$ is crossed the solution is non-singular. In all cycles as the Universe contracts below a certain size, and before it expands beyond the same size, there is an Euclidean phase. There is also a crossing of $\alpha=0$ in the Lorentzian phase. Thus we have eternal oscillations around $\alpha=0$ up to a maximal $|\alpha|$, with an oscillation between Euclidian and Lorentzian signatures in each cycle.

\item In Case 3, ${\cal V}^{2}$ asymptotes to the same, constant positive value of ${\cal V}^{2}$ as proper time tends to $-\infty$ and $+\infty$, thus once again asymptotically recovering General Relativity with identical, positive cosmological constant in each limit. In between these limits, as illustrated in Figure 4, ${\cal V}^{2}$ transitions to a negative value; during this period of negative ${\cal V}^{2}$,  $\alpha$ passes through zero (without singularity). Thus, we have contraction from $\alpha=-\infty$ to $\alpha=0$ followed by expansion to  $\alpha=\infty$, with an Euclidian phase around $\alpha=0$, and Lorentzian Einstein gravity asymptotically. The value of the cosmological constant $\Lambda$ asymptotically is positive and identical in each case.

\item In Case 4, ${\cal V}^{2}$ asymptotes to the same, positive constant value of ${\cal V}^{2}$. Although varying in between these limits, ${\cal V}^{2}$ never changes sign, and so the Universe is always Lorentzian. Unlike Case $1$,  this case involves $\alpha$ passing through zero. Like Case 3,  we have contraction from $\alpha=-\infty$ to $\alpha=0$ followed by expansion to  $\alpha=\infty$, but without signature change.

\item Case 5 may be seen as an Euclidean mirror-image of Case 4. ${\cal V}^{2}$ asymptotes to the same, negative constant value of ${\cal V}^{2}$ and does not change sign in between these limits. As in case 4, $\alpha$ passes through 0. By considering the analog of the action \eqref{4action} it would be found that asymptotically it is Euclidean General Relativity with a negative cosmological constant that is recovered. The geometry of such a solution in General Relativity is that of a surface $-w^{2}+x^{2}+y^{2}+z^{2}+v^{2} = -\mathrm{const}^{2}$ embedded in a five-dimensional Minkowski space of signature $(-,+,+,+,+)$ i.e. the surface is a higher dimensional hyperboloid of two-sheets. The intervening modification to General Relativity may be seen as a Euclidean `bridge' that joins surfaces together that asymptote to the above two sheets.

\item Case 6 may be seen as the mirror-image of Case 3. Now the Universe is Euclidean either side of a Lorentzian phase around $\alpha=0$. The Euclidean regimes asymptote to Euclidean General Relativity with a negative cosmological constant as in Case 5.

\item The cases $7a)$ and $7b)$ may be seen as the Euclidean mirror-image of cases $1a)$ and $1b)$.
\end{enumerate}
It is noteworthy that the majority of these cases asymptote to General Relativity (either Lorentzian or Euclidean) with a cosmological constant that gives rise to a scale factor varying exponentially with respect to the parameter $\tau$.. In Figure 5 we pick two of these cases to illustrate the point made at the end of Section 5, regarding the different possible metrics that can be adopted. We see that for the solution which contains a Euclidean regime between asymptotically Lorentzian General Relativity regimes, the sign of ${\cal G}_{\tau\tau}$ and $g_{\tau\tau}$ do not always agree but agree at large $|\tau|$; for the solution with oscillating signature of the metric $g_{\mu\nu}$, we note that the sign of ${\cal G}_{\tau\tau}$ and $g_{\tau\tau}$ always agree. 

Finally we can consider the nature of the solutions with $V^{a}(T_{0})=0$ and/or $W^{a}(T_{0})=0$. By inspection the contribution of the $b_{2}$ term to the equations of motion completely vanishes in this limit and so for $T\rightarrow T_{0}$ we expect solutions to asymptote to the corresponding case when only $a_{1}\neq 0$.  Consequently for $T$ close to $T_{0}$,  $V^{2}$ will evolve as in (\ref{fn}), and in doing so its evolution will become sensitive to the influence of terms in the equations of motion due to the $b_{2}$ term. It is conceivable that regimes exist in the early universe with $V^a=0$, so that the  $SO(1,1)$ symmetry is unbroken by the Higgs field. If that does happen, with a phase transition leading to broken symmetry, it is interesting to speculate whether there may be remnant topological defects corresponding to, for instance, signature change surfaces.

\begin{figure}
\centering
\includegraphics[width=15cm]{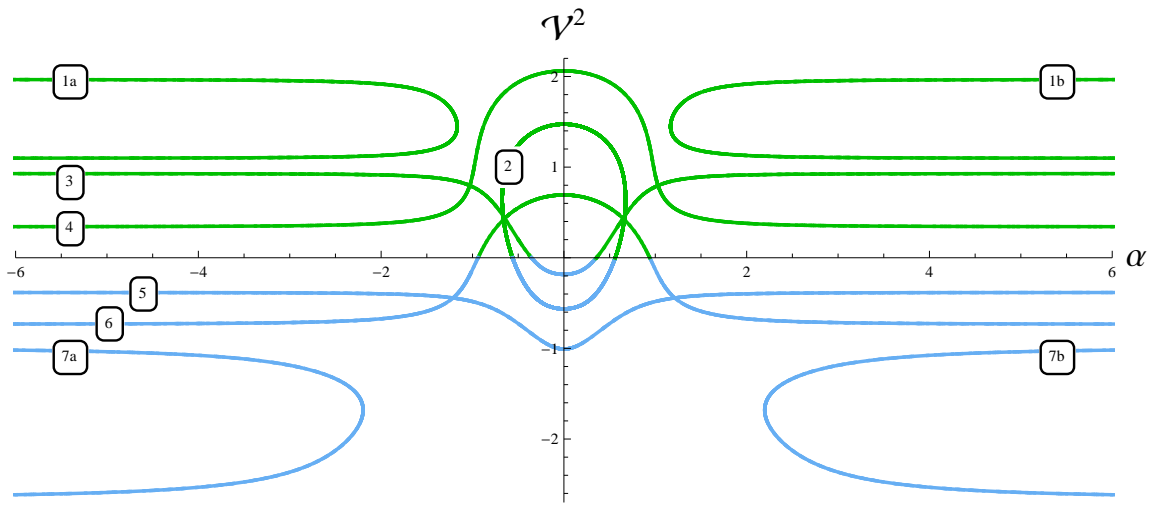} 
\label{cases2}
\caption{{\small Parametric plot displaying the dimensionless scale-factor as a function of proper-time, ${\cal \alpha}(\tau)$, and the $SO(1,4)$ norm of the gravitational Higgs field as a function of proper time, ${\cal V}^{2}(\tau),$ for the system where $a_{1}$ and $b_{2}$ are non-zero. We have labelled the various qualitatively different cases, as referred to in the main text. Lines and sections of line in green represent Lorentzian signature, those in blue denote Euclidean signature.}}
\end{figure} 

\begin{figure}
\centering
\includegraphics[width=10cm]{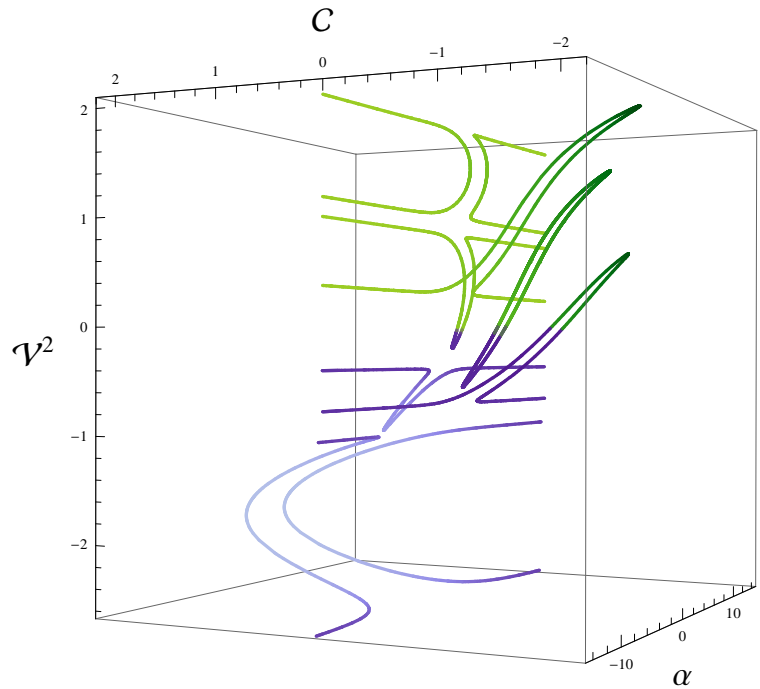} 
\caption{{\small Three dimensional parametric plot displaying solutions in terms of ${\cal \alpha}(\tau)$, ${\cal V}^{2}(\tau)$ and ${\cal C}(\tau)$. We have plotted the same cases which appear in Figure 2, but dropped the labels for clarity. As in Figure 2, green sections represent Lorentzian signature, blue denotes Euclidean signature.} }
\label{3d}
\end{figure}

\begin{figure}
\label{curve}
\centering
\includegraphics[width=12cm]{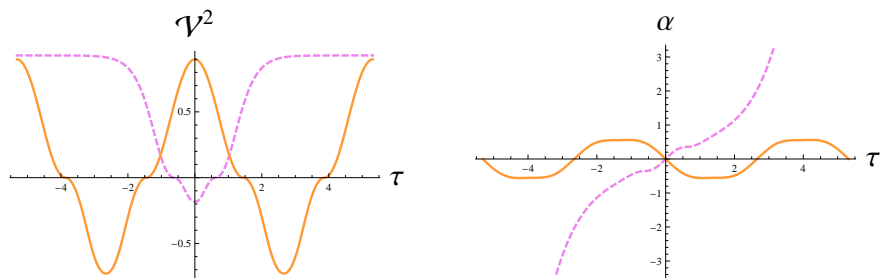} 
\caption{{\small The evolution of ${\cal V}^{2}$ and $\alpha$ as a function of proper-time $\tau$ for a solution containing a Euclidean region between asymptotically Lorentzian General Relativity (dashed line) and a solution with oscillating sign of ${\cal V}^{2}$ (solid).}}
\end{figure} 

\begin{figure}
\label{curve2}
\centering
\includegraphics[width=12cm]{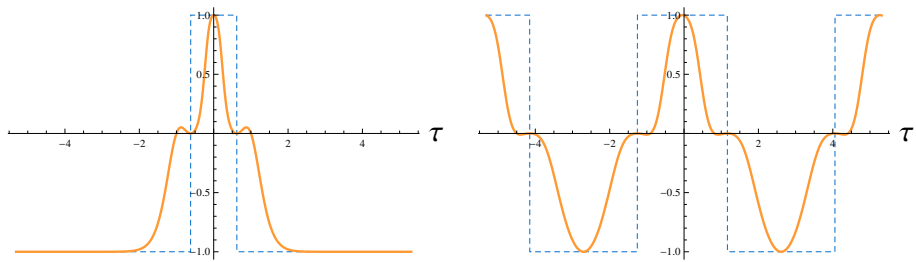} 
\caption{{\small Plot of ${\cal G}_{\tau\tau}$ (solid line) vs $g_{\tau\tau}$ (dashed) as a function of parameter  ${\tau}$ for asymptotically Lorentzian General Relativity solution (left) and oscillating signature solution (right)}.}
\end{figure}

\section{More general actions and quintessence}
\label{b1b2}

Thus far we have not considered the effects of the $\{a_{2},a_{3},b_{1},b_{3},c_{1}\}$. In Appendix \ref{genaction}, the action for arbitrary, constant $\{a_{i},b_{i},c_{1}\}$ in FRW symmetry is presented up to boundary terms.  Some general aspects of the influence of these terms were discussed in \cite{Westman:2013mf} without any particular spacetime symmetry assumed. By way of simplification, it may be shown that for constant $a_{3}$, the accompanying contribution to the Lagrangian is a boundary term and so will not contribute classically to the dynamics. Furthermore it may be shown via integration by parts that a constant $a_{2}$ term contributes identically to the equations of motion as does the term $b_{3}$. It is notable that a particular sub-case of the action (\ref{ca}) corresponds to the widely studied Peebles-Ratra quintessence model; this is the case where only $b_{1}$ and $b_{2}$ are non-zero. This is proved for a general Lorentzian spacetime (i.e. a spacetime where we may assume $V^{2}>0$) in \cite{Westman:2013mf}. Considered as a specific case of (\ref{act1}) and assuming that $b_{1}$ and $b_{2}$ are constant we have

\begin{eqnarray}
S_{b_{1}b_{2}}[\phi,e^{I},\omega^{IJ}] &=&  \int  b_{1}\phi\epsilon_{IJKL}\left(e^{I}e^{J}R^{KL}-\frac{1}{\phi^{2}}e^{I}e^{J}e^{K}e^{L}\right) +\frac{b_{2}\phi^{2}}{2}(T^{I}T_{I}-e_{I}e_{J}R^{IJ})\bigg) \label{actpr}
\end{eqnarray}
From the equations of motion obtained by varying with respect to $\omega^{IJ}$ we may solve for the contorsion $C^{IJ}$:

\begin{eqnarray}
C_{IJ} = \frac{1}{2\phi^{2}}e_{[I}\partial_{J]}\phi^{2}+\frac{b_{2}}{8b_{1}\phi}\epsilon_{IJKL}\partial^{K}\phi^{2}e^{L}
\end{eqnarray}
where $\partial^{L} \equiv e^{\mu L}\partial_{\mu}$. Insertion of this solution into the action (\ref{actpr}) we obtain an action that is a functional only of $\phi$ and $g_{\mu\nu}\equiv \eta_{IJ}e^{I}_{\mu}e^{J}_{\nu}$. Upon a conformal rescaling of $\tilde{g}_{\mu\nu}=\phi g_{\mu\nu}$ and redefinition of $\phi$ by a constant factor one recovers, up to boundary terms

\begin{eqnarray}
S'_{b_{1}b_{2}} &=& \int d^{4}\sqrt{-\tilde{g}}\left(\kappa_{1}\tilde{R}- \tilde{g}^{\mu\nu}\partial_{\mu}\phi\partial_{\nu}\phi -\frac {\kappa_{2}}{\phi^{3}}\right) \label{spr}.
\end{eqnarray}
This is an example of Peebles-Ratra quintessence \cite{Ratra:1987rm,Peebles:1987ek}, and it would be interesting to investigate how a signature change scenario could be integrated with a late-time acceleration period, and how they would interact. We have confirmed numerically that a system with non-zero $\{a_{1},b_{1},b_{2}\}$ can indeed exhibit an intermediate signature change regime between asymptotically tending to Peebles-Ratra quintessence as described by \eqref{spr}; however, we defer to a future publication a more comprehensive analysis of the full parameter space of these theories.

\section{Discussion}
In this paper we have examined extensions of General Relativity which are 
natural from the point of view of the first order, or 
``Einstein-Cartan'' formalism,
but not within the context of the second order formalism. 
The idea used for modifying the dynamics
is similar in flavour to the compactification of extra spacetime dimensions, 
but instead it is based on the introduction of a larger {\it internal} 
symmetry group, 
which is then broken (i.e. ``internally compactified'') 
to the usual Lorentz group. 
This is achieved by a mechanism reminiscent of the 
Englert-Brout-Higgs-Guralnik-Hagen-Kibble
mechanism for the electroweak interactions, and 
the idea was used before \cite{Stelle:1979va,Pagels:1983pq} 
to explain the awkward existence of the tetrad, beside the gauge field,
in the Einstein-Cartan formulation \cite{Westman:2012zk}. The glaring presence of a metric
field sets gravity apart from other other field theories in physics (though of course they couple to the metric). By extending the gauge group (for example, to the de Sitter group) 
and then spontaneously breaking it by means of a ``gravitational Higgs field'' $V^a$,
the tetrad emerges naturally. In the approaches of \cite{Stelle:1979va,Pagels:1983pq} 
the symmetry breaking field is non-dynamically forced to have a constant modulus (and be space-like). By dropping this restriction we are naturally led to an extension of Einstein-Cartan gravity. 
As we have demonstrated in this paper, many solutions are characterized by $V^2$ approaching a constant value for large proper times $|\tau|$ and so reduce to Einstein gravity with cosmological constant. That this is indeed possible is by no means trivial given the unfamiliar form of the polynomial action \eqref{ca}.

That $V^2$ settles down to a constant value would appear similar to symmetry breaking in the electroweak theory where $|\Phi|^2$ attains a constant value at sufficiently small energies. However, it should be stressed that the dynamical reasons for this behavior are distinct. In the electroweak theory the constancy of $|\Phi|^2$ is due to the Mexican hat shaped potential which is designed to have a specific minimum. In contrast, the approach to a constant $V^2$ in the $a_1-b_2$ action is not due to some Mexican hat shaped potential, not even in disguise. This can immediately be understood from the fact that different solutions to the same equations of motion can have different asymptotic values of $V^2$. This is not possible within the electroweak theory since the asymptotic value of $|\Phi|^2$ always coincides with the minimum valued of the Mexican hat potential. Thus, within this action is a new mechanism for achieving a constant value of $V^2$. By recasting the equation for $V^2$ into a second order form we see that a viscous term appears and this suggests that the constancy of $V^2$ is caused by `friction' causing the velocity $|d V^{2}/d\tau |$ to decrease. Note that this mechanism is entirely distinct from the mechanism discovered in \cite{Westman:2013mf} wherein it was found that certain combinations of $\{a_{i},b_{i},c_{i}\}$ terms in the action were equivalent to a scalar-tensor theory equipped with a potential with stable minimum at non-zero $V^{2}$. We leave it for future investigations to determine whether the former, new symmetry breaking mechanism also works outside the cosmological framework developed in this paper.

Indeed, the theory discussed in this paper is very general, and is represented by the action
(\ref{4action}). It turns out to be a theory with an Einstein-Cartan term
and a cosmological ``constant'', as well as Holst term, Euler, Pontryagin  
and Nieh-Yan boundary terms; however all these terms appear multiplied by 
factors that depend on a field $\phi$ (representing the modulus of $V^a$)
in a very specific form laid out in
Eqns.~(\ref{funcs}). Therefore
the usual ``boundary terms'' are no longer necessarily pure boundary terms. Newton's
``constant'' and Lambda are also typically functions of $\phi$. In fact, the cosmological 
term can never be independent of $\phi$ constant for actions polynomial in $\{A^{ab},V^{a}\}$, and specifically we recover the Peebles-Ratra
quintessence model. The field $\phi$ has propagating dynamics, even though 
this is not evident in the first-order action, and only becomes clear 
when we eliminate degrees of freedom, appealing to the torsion equation.

Remarkably this new theory allows for ``deterministic'' classical signature change
in the following sense. As explained in Section~\ref{sign}, it is possible to construct
solutions in the second order formalism which appear to change signature classically. Although such solutions exist, one may argue that they do not appear naturally within the standard metric formulation. To quote Ellis et al \cite{Ellis:1991st}:
\begin{quote}
`{\em The Einstein field equations by themselves do not determine the spacetime signature; that is imposed as an extra assumption}'.
\end{quote}
In contrast we see that in the theory envisaged here the signature changes has part of the ``deterministic'' classical dynamics of a gravitational Higgs field. The phenomenon occurs whenever the dynamics takes the field's $SO(1,4)$ norm, $V^aV_a$, from positive to negative, or vice versa. We have found a large array of such solutions, ranging from very simple to very complex, some more realistic than others. 

Specifically, we noted that the form of the functions multiplying the 
various terms in the action depends on the coefficients chosen for the unbroken
theory. In the simplest case we can turn-on only one of these terms,
the ``$a_1$-term''. A very simple analytical solution, exhibiting signature
change, was found in Section~\ref{a1}. Unfortunately
when we study solutions to this theory we found that it never becomes 
Einstein-Cartan asymptotically, i.e. the modulus of the symmetry breaking 
field never  stabilizes. We can regard it as a  useful toy-model for
signature change in modified gravity, but nonetheless were led to seek 
more complex, but more realistic solutions in Section~\ref{a1a2b2}, based on 
adding on more terms to the action (the ``$b_2$ term'', specifically).

In this context we found a large array of solutions, including some which
do asymptote to Einstein gravity when the universe is large, but experience
signature change when the universe is small, first in a contracting, then
in an expanding phase. We have also found other interesting oddities, such
as eternally oscillating universes, with the signature oscillating between 
Lorentzian and Euclidian. We find also many solutions without signature
change, both Euclidean and Lorentzian. In particular there are bouncing 
universes without signature change in this model. Whether or not these
classical solutions are realized, it is of note that they would have
to be included in any gravitational path integral.

In closing we mention a few open issues, left unresolved in this paper.
The coefficients $\{a_{i},b_{i},c_{i}\}$ (which could be promoted to functions of available de Sitter invariant scalars such as $V^{2}$) 
collectively amount to a vast parameter space. We have explored only
a small corner of this space, with interesting conclusions, but the question
arises as to whether these features are generic within these models, and
whether other types of behaviour exist. It is conceivable that more basic principles may ultimately place restrictions on the expected relative size of the $\{a_{i},b_{i},c_{i}\}$. By way of example, it is known that the Lagrangian $\epsilon_{abcde}V^{e}F^{ab}F^{cd}$ in isolation can arise following dimensional compactification of a five dimensional theory based on the Chern-Simons five-form for the group $SO(1,5)$ \cite{Chamseddine:1989nu}. 

In addition one can investigate the effects of matter coupling in the Cartan gravity 
description. The coupling of spinor, scalar, and gauge fields to the gravitational fields $\{A^{ab},V^{a}\}$
in the limit where the norm $V^{2}$ is fixed has been investigated \cite{Pagels:1983pq,Westman:2012zk} . Generalisation to the case of a truly dynamical $V^{2}$ remains; one may wonder whether, for instance, the presence of a matter scalar field prevent signature change from happening? Is there a deeper insight into what happens 
to scalar field dynamics in the presence of signature change in 
Cartan gravity vs metric General Relativity?
We hope to devote some work in the 
future to a more comprehensive exploration of these theories.

\vspace{1cm}

\section*{Acknowledgments} 

JM and TZ were funded by STFC through a consolidated grant. HW was supported by the Spanish
MICINN/MINECO Project FIS2011-29287, the CAM research consortium QITMAD S2009/ESP-1594, and the CSIC JAE-DOC 2011 program.

\appendix
\section{Imposing FRW symmetry}
\label{FRWSYM}
The {\em Cosmological Principle} dictates that the universe is spatially homogeneous and isotropic at large scales. Such symmetry is commonly referred to as FRW symmetry. From a mathematical point of view we require that our solutions are invariant under diffeomorphisms representing rotations and translations/transvections. If space is three-dimensional we have three rotations and three translations and thus our symmetry group should be six-dimensional, i.e. we have six Killing vectors which can be shown to take the form (see e.g. \cite{Toloza:2013wi})
\begin{align}
\xi_{(i)}=\sqrt{1-kr^2}\frac{\partial}{\partial x^i}\quad \xi_{(ij)}=x_i\frac{\partial}{\partial x^j}-x_j\frac{\partial}{\partial x^i}
\end{align}
where $k=-1,0,+1$ and $r^2=\delta_{ij}x^ix^j$ and the $x^{i}$ coordinates are related to spherical coordinates $(r,\theta,\varphi)$ in the manner that Cartesian coordinates are.  Indeed for our purposes it is more convenient to express these Killing vectors in spherical coordinates:
\begin{align}
\xi_{(12)}=\partial_\varphi\qquad \xi_{(31)}=\cos\varphi\partial_\theta-\frac{\sin\varphi}{\tan\theta}\partial_\varphi\qquad \xi_{(23)}=-\sin\varphi \partial_\theta-\frac{\cos\varphi}{\tan\theta}\partial_\varphi.
\end{align}
The different values of $k$ correspond to the only three possible groups that are compatible with homogeneity and isotropy:
\begin{itemize}
\item $k=0$: the commutators of these six Killing vectors satisfy the Lie-algebra of the inhomogeneous Euclidean group $ISO(3)$, i.e. the symmetry group of an infinite flat Euclidean space
\item $k=+1$: the commutators of these six Killing vectors satisfy the Lie-algebra $SO(4)$, i.e. the symmetry group of the three-sphere $S^3$
\item $k=-1$: the commutators of these six Killing vectors satisfy the Lie-algebra of $SO(1,3)$, i.e. the symmetry group of an infinite hyperbolic three-dimensional space.
\end{itemize}
To achieve homogeneity and isotropy in the metric formulation we would simply impose the conditions
\begin{align}
\m L_{\xi_{(i)}}g_{\mu\nu}=\m L_{\xi_{(ij)}}g_{\mu\nu}=0\nn
\end{align}
where $\m L_\xi$ is the Lie-derivative along a vector field $\xi$ and geometrically is understood as an infinitesimal diffeomorphism. However, Cartan gravity operates with different fundamental variables from that of the second order metric general relativity. Instead of a metric tensor $g_{\mu\nu}$ we have the two objects; a scalar $V^a$ and a connection $A^{ab}$ both valued in the Lie-algebra $\mathfrak{so}(1,4)$. The presence of non-tensor indices, i.e. the $SO(1,4)$ gauge indices $a$ and $b$, poses some challenges for how to impose FRW symmetry on our variables, i.e. homogeneity and isotropy. The reason for this is that any equation such as $\m L_{\xi_{(i)}}V^{a}=0$, as may be checked, is not gauge covariant.
One suitable approach is to require that all the possible $SO(1,4)$ invariant tensors built out of the pair $\{A^{ab},V^{a}\}$ should exhibit FRW symmetry. For example, in a open set where $V^2>0$ we can always gauge fix so that $V^a\overset{*}{=}\phi\delta^a_4$ and $e^I\overset{*}{=}DV^I$. This co-tetrad $e^I_\mu$ must yield a FRW symmetric metric $g_{\mu\nu}=\eta_{IJ}e^I_\mu e^J_\nu$ (see e.g. \cite{Wald:1984rg})
\begin{align}\label{FRWmetric}
g_{tt}=g_{tt}(t)\qquad g_{rr}=\frac{a^2(t)}{K(r)^2} \qquad g_{\theta\theta}=a^2(t)r^2\qquad g_{\varphi\varphi}=a^2(t)r^2\sin^{2}\theta \qquad K(r)=\sqrt{1-kr^2}.
\end{align}
A convenient choice of co-tetrad $e_\mu^I$ that yields \eqref{FRWmetric} is
\begin{align}
e^0_t=\sqrt{|g_{tt}|}\qquad e^1=\frac{a(t)}{K(r)}dr\qquad e^2=a(t)rd\theta\qquad e^3=a(t)r\sin\theta d\varphi.
\end{align}
Furthermore, whenever the inverse $e^\mu_I$ of $e^I_\mu$ exists the torsion tensor
\begin{align}
T_{\mu\nu}^{\ph{\mu\nu}\rho}\equiv e^\rho_IT_{\mu\nu}^{\ph{\mu\nu}I}
\end{align}
must also display FRW symmetry, i.e. satisfy $\m L_{\xi_{(i)}}T_{\mu\nu}^{\ph{\mu\nu}\rho}=\m L_{\xi_{(ij)}}T_{\mu\nu}^{\ph{\mu\nu}\rho}=0$ which yields the most general functional form \cite{Toloza:2013wi}
\begin{align}
T_{\theta\varphi}^{\ph{\theta\varphi}r}&=f(t)r^2K(r)\sin\theta\qquad T_{r\theta}^{\ph{r\theta}\varphi}=\frac{f(t)}{K(r)\sin\theta}\nn\\
T_{r\varphi}^{\ph{r\varphi}\theta}&=-\frac{f(t)\sin\theta}{K(r)}\qquad T_{tr}^{\ph{tr}r}=T_{t\theta}^{\ph{t\theta}\theta}=T_{t\varphi }^{\ph{t\varphi}\varphi}=g(t) 
\end{align}
or using $T^I=\frac{1}{2}e^I_\rho T_{\mu\nu}^{\ph{\mu\nu}\rho}dx^\mu dx^\nu$
\begin{eqnarray}
T^{i} &=& g(t) e^{i}e^{0}+f(t)\epsilon^{i}_{\phantom{i}jk}e^{j}e^{k}
\end{eqnarray}

It may further be checked that $T^{0}=0$..
Using the definition of the torsion two-form $T^I=de^I+\omega^{IJ}e_J$ allows us to read off the most general functional form of the spin connection $\omega^{IJ}$
\begin{align}
\omega^{0i}&=B(t)E^i\qquad \omega^{12}=-\frac{K(r)}{r}E^2-C(t)E^3\nn\\
\omega^{13}&=-\frac{K(r)}{r}E^3+C(t)E^2\qquad \omega^{23}=-\frac{\cot\theta}{r}E^3-C(t)E^1.
\end{align}
In the gauge $V^a\overset{*}{=}\phi \delta^a_4$ we can now deduce the most general functional form of the $SO(1,4)$ connection
\begin{eqnarray}
A^{ab}&\overset{*}{=}&\left(\begin{array}{ccc}0&B(t)E^j&N(t)E^0\\-B(t)E^i&\omega^{ij}&A(t)E^i\\-N(t)E^0&-A(t)E^j&0\end{array}\right). \label{aee}
\end{eqnarray}
We stress that this functional form of $A^{ab}$ was obtained under the assumption that $V^a$ is space-like. However, it is straightforward to verify that starting with a time-like $V^a$ yields the same functional form for $A^{ab}$. Therefore \eqref{aee} is the most general FRW symmetric form of $A^{ab}$.

From the gravitational Higgs field $V^a$  we can form the gauge invariant scalar $V^2=\eta_{ab}V^aV^b$. Imposing FRW symmetry yield $V^2=V^2(t)$. Secondly, we can always adopt a gauge such that $V^i\overset{*}{=}0$. Therefore the form of the gravitational Higgs field is 
\begin{align}
V^a\overset{*}{=}(\psi(t),0,0,0,\phi(t)).
\end{align}
\section{General Action}
\label{genaction}
Imposing FRW symmetry on the action \eqref{ca} yields, up to boundary terms, 
\begin{eqnarray}
S= \int L dt &\equiv & \int \left(L_{a_{1}}+L_{a_{2}}+L_{b_{1}}+L_{b_{2}}+L_{c_{1}}\right)dt
\end{eqnarray}
where
\begin{eqnarray}
L_{a_{1}} &=& 4a_{1}\left(-\epsilon_{AB}V^{A}\m DW^{B}\left(k-W^{2}-C^{2}\right)-2\epsilon_{AB}W^{A}V^{B}C\m DC\right) \nonumber\\
 &=& 4a_{1}\epsilon_{AB}\left(V^{B}\m DW^{A}\left(k-W^{2}\right)+C^{2}W^{A}\m DV^{B}\right)\\
L_{a_{2}} &=& 2a_{2}CW_{A}V^{A}V_{B}\m DW^{B}\\
L_{b_{1}} &=& 2b_{1}\left(W_{A}V^{A}\epsilon_{BC}V^{B}\m DV^{C}\left(k-W^{2}-C^{2}\right)-(W_{A}V^{A})^{2}\epsilon_{BC}V^{B}\m DW^{C}\right)\\
L_{b_{2}} &=& -\frac{b_{2}}{2}C(W_{A}V^{A})^{2}\m DV^{2} \\
L_{c_{1}} &=& 4c_{1} (W_{A}V^{A})^{3}\epsilon_{BC}V^{B}\m DV^{C}.
\end{eqnarray}
The contribution due to the $b_{3}$ term is largely similar to the contribution to the $a_{2}$ term.

\section{Derivation of $a_{1}b_{2}$ equations of motion}
\label{devab}

The $a_1-b_2$ action is given by
\begin{eqnarray}
S_{a1b2}
&=& \int dt4a _1\left(\bar V_A\m DW^A(k-W^2)+C^2W^A\m D\bar V_A-\chi CV_A\m D V^A (W^BV_B)^2\right)\nn
\end{eqnarray}
which yields the equations of motion
\begin{align}
N&:\qquad 0=V\cdot W(k-W^2-C^2)\\
W^A&:\qquad 0=-\m D(\bar V_A(k-W^2-C^2))-2\bar V_B\m DW^B W_A-2\dot CC\bar V_A-\frac{\chi}{2} V_A C\m DV^2 (W\cdot V)\\
V^A&:\qquad 0=-\m D\bar W_A(k-W^2-C^2)+2\dot CC\bar W_A+\chi V_A \m D(C(W\cdot V)^2)-\chi C\m D V^2 W_A(W\cdot V)\\
C&:\qquad 0=2CW^A\m D\bar V_A-\frac{\chi}{2}\m DV^2 (W\cdot V)^2.
\end{align}
Adopting the solution $0=k-W^{2}-C^{2}$ to the $N$ equation of motion and implementing this restriction in the remaining equations we have
\begin{eqnarray}\label{ze1}
W^A&:&\qquad 0=2\dot CC\bar V_A+2\bar V_B\m DW^B W_A+\frac{\chi}{2} V_A C\m DV^2 (W\cdot V)\\
V^A&:&\qquad 0=2\dot CC\bar W_A+\chi V_A \m D(C(W\cdot V)^2)-\chi W_A C\m D V^2 (W\cdot V)\\
C&:&\qquad 0=2CW^A\m D\bar V_A-\frac{\chi}{2}\m DV^2 (W\cdot V)^2. \label{ze2}
\end{eqnarray}
Let us then make the following ansatz for $W^{A}$:
\begin{align}
W^A=\frac{1}{2}(k-C^2+1)\m W^A+\frac{1}{2}(k-C^2-1)\bar{\m W}^A
\end{align}
where ${\cal W}^{A}{\cal W}_{A}=1$. As may be checked we have
\begin{align}
W^2=\frac{1}{4}\left((k-C^2+1)^2\m W^2+(k-C^2-1)^2\bar{\m W}^2\right)=\frac{1}{4}\left((k-C^2+1)^2-(k-C^2-1)^2\right)=k-C^2
\end{align}
and we also have
\begin{align}
DW^A=\frac{1}{2}(k-C^2+1)D\m W^A+\frac{1}{2}(k-C^2-1)D\bar{\m W}^A-\dot CC(\m W^A+\bar{\m W}^A)=\m N \bar W^A-\dot CC(\m W^A+\bar{\m W}^A)
\end{align}
which yields
\begin{align}
\bar V_ADW^A&=\bar V_A(\m N \bar W^A-\dot CC(\m W^A+\bar{\m W}^A))=\m N a-\dot CC(\bar V\cdot\m W-V\cdot\m W)\\
a&=-V_AW^A=\frac{1}{2}(k-C^2-1)\bar V\cdot\m W-\frac{1}{2}(k-C^2+1)V\cdot\m W \label{ce1}\\
\bar a&=\bar V_A W^A=\frac{1}{2}(k-C^2+1)\bar V\cdot\m W-\frac{1}{2}(k-C^2-1)V\cdot\m W. \label{ce2}
\end{align}
Furthermore using equations (\ref{ce1}) and (\ref{ce2}) we have that 
\begin{align}
\bar a+a&=\frac{1}{2}(k-C^2-1)\bar V\cdot\m W-\frac{1}{2}(k-C^2+1)V\cdot\m W+\frac{1}{2}(k-C^2+1)\bar V\cdot\m W-\frac{1}{2}(k-C^2-1)V\cdot\m W\nn\\
&=(k-C^2)(\bar V\cdot\m W-V\cdot\m W)
\end{align}
so that
\begin{align}
\bar V_ADW^A=\m N a-\dot CC(\bar V\cdot\m W-V\cdot\m W)=\m N a-\dot CC\frac{\bar a+a}{k-C^2}.
\end{align}
Using the above expressions in the equations of motion (\ref{ze1}) to (\ref{ze2}) and contracting the equations variously with $W^{A}$ with $\bar W^{A}$ we have
\begin{align}
0&=(k-C^2)CDV^2-\dot Ca^2-2C\dot aa \label{dee1}\\
0&=-\dot CC(k-C^2)-(\chi \dot Ca^2+2\chi C\dot aa)\bar a\\
0&=-\left(\m N a-\dot CC\frac{\bar a+a}{k-C^2}\right)(k-C^2)-\dot CC\bar a-\chi aCaDV^2\\
\dot C&=-\chi \bar aDV^2\\
0&=-2C\left(\m N a-\dot CC\frac{\bar a+a}{k-C^2}\right)+2C\dot{\bar a}-\chi a^2DV^2. \label{dee2}
\end{align}
Thus we see we have written the evolution equations for this system in terms of time derivatives of gauge-invariant quantities $\{a,\bar{a},C,V^{2}\}$. Multiplying the constraint condition $0=k-W^{2}-C^{2}$ by $V^{2}$ we obtain the constraint $a^2-\bar a^2=(k-C^2)V^2=(a-\bar a)(a+\bar a)$, which may be used to eliminate $k$ from equations (\ref{dee1}) and (\ref{dee2}):
\begin{align}
\dot a&=-\m N\frac{a-\bar a}{2\chi C^2V^2}\frac{C(k-C^2)+\chi \bar aa^2}{a}\\
DV^2&=-\m N2C\frac{a-\bar a}{2\chi C^2V^2}\\
\dot C&=\m N2\chi\bar aC\frac{a-\bar a}{2\chi C^2V^2}\\
\dot{\bar a}&=\m N\frac{a-\bar a}{2\chi C^2V^2}\chi(2C^2V^2-a^2).
\end{align}

\bibliographystyle{hunsrt}
\bibliography{references}

\end{document}